\documentclass[stslayout, noinfoline]{imsart}
\usepackage{geometry} 
\usepackage{graphicx}	
\usepackage{amssymb, amsmath}
\usepackage{natbib}
\usepackage[colorlinks,citecolor=blue,urlcolor=blue]{hyperref}
\usepackage{braket}

\usepackage{subfigure}
\usepackage{verbatimbox}
\usepackage{tikz}
\usetikzlibrary{calc, arrows.meta}
\usepackage{pgfmath}

\definecolor{light}{RGB}{220, 188, 188}
\definecolor{mid}{RGB}{185, 124, 124}
\definecolor{dark}{RGB}{143, 39, 39}
\definecolor{highlight}{RGB}{0, 255, 0}
\definecolor{gray10}{gray}{0.1}
\definecolor{gray20}{gray}{0.2}
\definecolor{gray30}{gray}{0.3}
\definecolor{gray40}{gray}{0.4}
\definecolor{gray60}{gray}{0.6}
\definecolor{gray70}{gray}{0.7}
\definecolor{gray80}{gray}{0.8}
\definecolor{gray90}{gray}{0.9}
\definecolor{gray90}{gray}{0.9}
\definecolor{gray91}{gray}{0.91}
\definecolor{gray92}{gray}{0.92}
\definecolor{gray93}{gray}{0.93}
\definecolor{gray94}{gray}{0.94}
\definecolor{gray95}{gray}{0.95}

\usepackage{systeme}

\usepackage{amsthm}

\theoremstyle{definition}

\usepackage{algorithm}
\usepackage[noend]{algpseudocode}

\begin{document}

\begin{frontmatter}

\title{Efficient Automatic Differentiation of Implicit Functions}
\runtitle{Efficient Automatic Differentiation of Implicit Functions}

\begin{aug}
\author{Charles C. Margossian and Michael Betancourt}
  \runauthor{Margossian and Betancourt}
  \address{Charles Margossian is a PhD candidate in the Department of Statistics, 
           Columbia University. Michael Betancourt is the principal research scientist 
           at Symplectomorphic, LLC.}
\end{aug}

\begin{abstract}
Derivative-based algorithms are ubiquitous in statistics, machine learning, and applied 
mathematics.  Automatic differentiation offers an algorithmic way to efficiently evaluate 
these derivatives from computer programs that execute relevant functions.  Implementing 
automatic differentiation for programs that incorporate implicit functions, such as the 
solution to an algebraic or differential equation, however, requires particular care.
Contemporary applications typically appeal to either the application of the implicit 
function theorem or, in certain circumstances, specialized adjoint methods.  In this paper 
we show that both of these approaches can be generalized to \textit{any} implicit function, 
although the generalized adjoint method is typically more effective for automatic 
differentiation.  To showcase the relative advantages and limitations of the two methods 
we demonstrate their application on a suite of common implicit functions.
\end{abstract}

\end{frontmatter}

\pagebreak

\tableofcontents

\pagebreak

Automatic differentiation is a powerful tool for algorithmically evaluating derivatives
of functions implemented as computer programs \citep{Griewank:2008, Baydin:2018, Margossian:2019}.
The method is implemented in an increasing diversity of software packages such as Stan 
\citep{Carpenter:2015, Carpenter:2017} and Jax \citep{Jax:2018}, driving state of the 
art computational tools such as the aforementioned Stan and TensorFlow \citep{Dillon:2017}.
Each automatic differentiation package provides a library of differentiable expressions 
and routines that propagate derivatives through programs comprised of those expressions.

Implicit functions are defined not as explicit expressions but rather by a potentially 
infinite set of equality constraints that an output must satisfy for a given input.  Common 
examples include algebraic equations, optima, and differential equations.  Although defined
only implicitly by the constraints they must satisfy, these functions and their derivatives
can be evaluated at a given input which puts them within the scope of automatic
differentiation.

Many approaches to evaluating the derivatives of implicit functions have been developed, 
most designed for specific classes of implicit functions.  Here we focus on two approaches:
direct application of the implicit function theorem and adjoint methods.  The former is
commonly applied to finite-dimensional systems such as algebraic equations and optimization 
problems \cite[for example][]{Bell:2008, Lorraine:2019, Gaebler:2021} while the latter is 
particularly well-suited to infinite-dimensional systems such as ordinary differential equations 
\cite[for example][]{Pontryagin:1963, Errico:1997}, algebraic differential equations 
\citep{Cao:2002}, and stochastic differential equations \citep{Li:2020}.  Adjoint methods 
have also been derived for some finite-dimensional systems such as difference equations 
\citep{Betancourt:2020}.  When they can be derived the performance of adjoint methods often 
scales better than the performance of implicit function theorem methods; the details
of those derivations, however, can change drastically from one system to another.

In this paper we derive implicit function theorem and adjoint methods that implement
automatic differentiation for \emph{any} implicit function regardless of its dimensionality.  
We begin by reviewing derivatives of real-valued functions -- drawing a careful distinction 
between total, partial, and directional derivatives -- and then introduce the basics of 
automatic differentiation.  Next we derive implicit function theorem and adjoint 
methods for any finite-dimensional implicit function and demonstrate their application 
to the reverse mode automatic differentiation of general algebraic equations, the special 
case of difference equations, and optimization problems.  Finally we generalize these methods to 
infinite-dimensional implicit functions with demonstrations on ordinary and algebraic 
differential equations.  In each example we examine the particular challenges that 
arise with each method.

\section{Automatic Differentiation}
Before discussing the automatic differentiation of implicit functions, in this section we will review 
the basics of differentiating real-valued functions and then how derivatives 
of computer programs can be implemented algorithmically as automatic 
differentiation.

\subsection{A Little Derivative}

Differentiation is a pervasive topic, but terminology and notation can vary 
strongly from field to field.  In this section we review the mathematics of 
derivatives on real spaces and introduce all of the terminology and notation 
that we will use throughout the paper.  We first discuss the parameterization of 
real spaces and the vector space interpretation that emerges before introducing 
formal definitions for total and directional derivatives and their properties.

\subsubsection{Locations and Directions.}

An $I$-dimensional real space $\mathbb{R}^{I}$ models a rigid and smooth 
continuum of points.  Here we will consider a subset of the real numbers 
$X \subseteq \mathbb{R}^{I}$ which may or may not be compact.

\begin{figure*}
\centering
\begin{tikzpicture}[scale=0.2, thick]

  \begin{scope}[shift={(0, 0)}]
    
    \begin{scope}
      \clip (-10, -10) rectangle (20, 20);
      
      \foreach \delta in {-8, -6, ..., 8} {
        \draw[gray80] (\delta, -10) -- (\delta, 10);
        \draw[gray80] (-10, \delta) -- (10, \delta);
      }
      
      \draw[black] (0, -10) -- (0, 10);
      \draw[black] (-10, 0) -- (10, 0);
      
    \end{scope}
    
    \node at (12, 0)  { $x_{1}$ };
    \node at (0, -12) { $x_{2}$ };
    
  \end{scope}
  
  \begin{scope}[shift={(25, 0)}]
    
    \begin{scope}
      \clip (-10, -10) rectangle (20, 20);
      
      \foreach \delta in {2, 4, ..., 20} {
        \draw[gray80] (-10, 10 - \delta) -- (10 - \delta, -10);
        \draw[gray80] (-10 + \delta, 10) -- (10, -10 + \delta);
        
        \draw[gray80] (-10, -10 + \delta) -- (10 - \delta, 10);
        \draw[gray80] (-10 + \delta, -10) -- (10, 10 - \delta);
      }
      
      \draw[black] (-10, -10) -- (10, 10);
      \draw[black] (-10, 10) -- (10, -10);
      
    \end{scope}
    
    \node at (12, 12)  { $x_{1}$ };
    \node at (12, -12) { $x_{2}$ };
    
  \end{scope}
  
  \begin{scope}[shift={(50, 0)}]
    
    \begin{scope}
      \clip (-10, -10) rectangle (10, 10);
      
      \foreach \delta in {-14, -12, ..., 14} {
        \draw[domain={-10:10}, smooth, samples=10, variable=\x, color=gray80] 
          plot ({\x}, {0.01 * \x * \x * \x + \delta});
        \draw[domain={-10:10}, smooth, samples=10, variable=\x, color=gray80] 
          plot ({0.01 * \x * \x * \x + \delta}, {-\x});
      }
      
      \draw[domain={-10:10}, smooth, samples=20, variable=\x, color=black] 
        plot ({\x}, {0.01 * \x * \x * \x});

      \draw[domain={-10:10}, smooth, samples=20, variable=\x, color=black] 
        plot ({0.01 * \x * \x * \x}, {-\x});
      
    \end{scope}
    
    \node at (12, 12)  { $x_{1}$ };
    \node at (12, -12) { $x_{2}$ };
    
  \end{scope}

\end{tikzpicture}
\caption{
Every real space $X$ admits an infinite number of parameterizations, or 
coordinate systems, each of which are capable of uniquely identifying 
every point with an ordered tuple of real numbers.
}
\label{fig:coordinate_systems} 
\end{figure*}
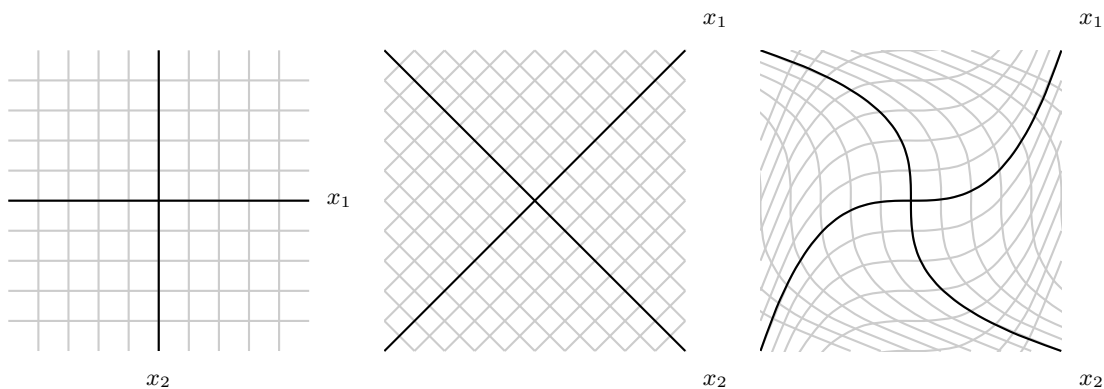

A \emph{parameterization} of $X$ decomposes the $I$-dimensional space into $I$ 
copies of the one-dimensional real line,
\begin{equation*}
X \approx \mathbb{R}_{1} \times \ldots \times \mathbb{R}_{i} \times \ldots \times \mathbb{R}_{I},
\end{equation*}
which we refer to as a \emph{coordinate system}.  Within a parameterization each point 
$x \in X$ can be identified by $I$ real numbers denoted \emph{parameters} or 
\emph{coordinates},
\begin{equation*}
x = (x_{1}, \ldots, x_{i}, \ldots, x_{I}).
\end{equation*}
Every real space $X$ admits an infinite number of parameterizations 
(Figure \ref{fig:coordinate_systems}).  A one-to-one map from $X$ into itself can be 
interpreted as a map from one parameterization to another and consequently is often 
denoted a \emph{reparameterization} or \emph{change of coordinate system}.

The choice of any given parameterization endows $X$ with a rich \emph{geometry}.
In particular we can use coordinates to define how to scale any point $x \in X$ 
by a real number $\alpha \in \mathbb{R}$,
\begin{align*}
\alpha \cdot x
&=
\alpha \cdot (x_{1}, \ldots, x_{i}, \ldots, x_{I})
\\
&=
(\alpha \cdot x_{1}, \ldots, \alpha \cdot x_{i}, \ldots, \alpha \cdot x_{I})
\\
&= x' \in X,
\end{align*}
as well as add two points $x, x' \in X$ together,
\begin{align*}
x + x' 
&=
(x_{1}, \ldots, x_{i}, \ldots, x_{I})
+
(x'_{1}, \ldots, x'_{i}, \ldots, x'_{I})
\\
&=
( x_{1} + x'_{1}, \ldots, x_{i} + x'_{i}, \ldots, x_{I} + x'_{I})
\\
&= x'' \in X.
\end{align*}
These properties make the parameterization of $X$ a \emph{vector space
over the real numbers}; each point $x \in X$ identifies a unique vector 
$\mathbf{x}$ and the coordinates define a distinguished vector space 
basis.  If we further use the coordinates to define an inner product,
\begin{equation*}
\left< \mathbf{x}, \mathbf{x}' \right> \equiv \sum_{i = 1}^{I} (x_{i} - x'_{i})^{2}
\end{equation*}
then this vector space becomes a \emph{Euclidean vector space}, $E(X)$.

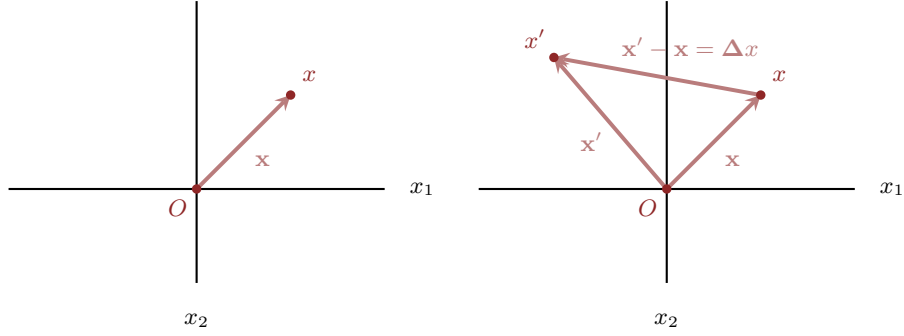
\begin{figure*}
\centering
\begin{tikzpicture}[scale=0.25, thick]

  \begin{scope}[shift={(0, 0)}]
    
    \draw[black] (-10, 0) -- (10, 0);
    \node at (12, 0)  { $x_{1}$ };  

    \draw[black] (0, -5) -- (0, 10);  
    \node at (0, -7) { $x_{2}$ };
 
    \draw[->, >=stealth, mid, line width=1.5] (0, 0) -- (5, 5);
    \node[mid] at (3.5, 1.5) { $\mathbf{x}$ };
    
    \fill[dark] (5, 5) circle (0.25);
    \node[dark] at (6, 6) { $x$ };

    \fill[dark] (0, 0) circle (0.25);
    \node[dark] at (-1, -1) { $O$ };
    
  \end{scope}

  \begin{scope}[shift={(25, 0)}]
    
    \draw[black] (-10, 0) -- (10, 0);
    \node at (12, 0)  { $x_{1}$ };  

    \draw[black] (0, -5) -- (0, 10);  
    \node at (0, -7) { $x_{2}$ };
    
    \draw[->, >=stealth, mid, line width=1.5] (0, 0) -- (5, 5);
    \node[mid] at (3.5, 1.5) { $\mathbf{x}$ };

    \draw[->, >=stealth, mid, line width=1.5] (0, 0) -- (-6, 7);
    \node[mid] at (-4, 2.5) { $\mathbf{x}'$ };
    
    \draw[->, >=stealth, mid, line width=1.5] (5, 5) -- (-6, 7);
    \node[mid] at (1.25, 7.5) { $\mathbf{x}' - \mathbf{x} = \boldsymbol{\Delta} x$ };
    
    \fill[dark] (5, 5) circle (0.25);
    \node[dark] at (6, 6) { $x$ };

    \fill[dark] (-6, 7) circle (0.25);
    \node[dark] at (-7, 8) { $x'$ };

    \fill[dark] (0, 0) circle (0.25);
    \node[dark] at (-1, -1) { $O$ };
    
  \end{scope}
  
\end{tikzpicture}
\caption{
Given a parameterization of the real space $X$ vectors quantify the direction 
and distance between points, such as $\mathbf{x}$ between point $x$ and the 
origin $O$ or $\mathbf{x}' - \mathbf{x}$ between $x$ and $x'$.  By following 
a vector we can also translate from the initial point to the final point.
}
\label{fig:translation} 
\end{figure*}

This Euclidean vector space structure defines a notion of \emph{direction}
and \emph{orientation} in $X$.  For example any point $x \in X$ can be 
interpreted as a vector $\mathbf{x}$ stretching from the origin at
$(0, \ldots, 0, \ldots, 0) \equiv O$ to $x$.  Similarly any two points 
$x, x' \in X$ are connected by the vector
\begin{align*}
\boldsymbol{\Delta} \mathbf{x}
&=
\mathbf{x}' - \mathbf{x}
\\
&=
(x'_{1} - x_{1}, \ldots, x'_{i} - x_{i}, \ldots, x'_{I} - x_{I}).
\end{align*}
Equivalently we can think of vectors as a way to translate from one point to 
another (Figure \ref{fig:translation}).  For example $\mathbf{x}$ shifts the 
origin to $x$,
\begin{align*}
x 
&= (x_{1}, \ldots, x_{i}, \ldots, x_{I})
\\
&= (0 + x_{1}, \ldots, 0 + x_{i}, \ldots, 0 + x_{I})
\\
&=
(0, \ldots, 0, \ldots, 0) +  (x_{1}, \ldots, x_{i}, \ldots, x_{I})
\\
&=
O + \mathbf{x},
\end{align*}
while $\boldsymbol{\Delta} \mathbf{x}$ shifts from $x$ to $x'$,
\begin{equation*}
x' = x + (x' - x) = x + (\mathbf{x}' - \mathbf{x}) = x + \boldsymbol{\Delta} \mathbf{x}.
\end{equation*}

Consequently once we fix a parameterization each set of coordinates 
$(x_{1}, \ldots, x_{i}, \ldots, x_{I})$ can be interpreted as either a 
\emph{location} $x \in X$ or a \emph{direction} $\mathbf{x} \in E(X)$.

\subsubsection{Transforming Locations and Directions.}

Given two real spaces $X \subseteq \mathbb{R}^{I}$ and $Y \subseteq \mathbb{R}^{J}$
a \emph{real-valued function} $f : X \rightarrow Y$ maps points in $X$ to points in $Y$.  
Once a parameterization has been fixed for both spaces such a mapping also induces a 
map from vectors in $E(X)$ to vectors in $E(Y)$, $F : E(X) \rightarrow E(Y)$.

Induced maps that preserve the additive and multiplicative structure of the 
Euclidean vector space,
\begin{equation*}
F(\alpha \cdot \mathbf{x} + \beta \cdot \mathbf{x}') 
= 
\alpha \cdot F(\mathbf{x}) + \beta \cdot F(\mathbf{x}'),
\end{equation*}
are said to be \emph{linear}.  Linear maps can be represented by a matrix 
of real numbers that maps the components of the input vector to the components 
of the output vector,
\begin{equation*}
y_{j} = \sum_{i = 1}^{I} F_{ji} \, x_{i},
\end{equation*}
or in standard linear algebra notation,
\begin{equation*}
\mathbf{y} = F(\mathbf{x}) = \mathbf{F} \cdot \mathbf{x}.
\end{equation*}
We will denote the space of linear maps between $E(X)$ and $E(Y)$ as $L(X, Y)$.

\subsubsection{The Total Derivative.}

The \emph{total derivative} of a function $f : X \rightarrow Y$ quantifies the 
behavior of $f$ in the local neighborhood of an input point $x \in X$.  This local 
behavior provides a way of mapping vectors that represent infinitesimal 
translations from $x$ to vectors that represent infinitesimal translations from 
$f(x)$.  

More formally the total derivative assigns to each point $x \in X$ a linear 
transformation between $E(X)$ and $E(Y)$,
\begin{alignat*}{6}
\frac{\mathrm{d}}{\mathrm{d}x} :\; &X & &\rightarrow& \; &L(X, Y)&
\\
&x& &\mapsto& &\frac{\mathrm{d} f}{\mathrm{d} x}(x) \equiv J(x)&,
\end{alignat*}
that quantifies the first-order variation of $f$ in the neighborhood of each 
input $x$.  We can then say that \emph{the total derivative of $f$ at $x$} is 
the linear transformation $J(x) : E(X) \rightarrow E(Y)$.

In components the total derivative at a point $x \in X$ is specified by a matrix of 
partial derivative functions evaluated at $x$,
\begin{equation*}
J_{ji}(x) = \frac{\partial f_{j}}{\partial x_{i}} (x),
\end{equation*}
denoted the \emph{Jacobian}.  The action of the total derivative at $x$ on a vector 
$\mathbf{v} \in E(X)$ is then given by matrix multiplication,
\begin{equation*}
J(x)(\mathbf{v}) = \mathbf{J}(x) \cdot \mathbf{v} =  \sum_{i = 1}^{I} J_{ji}(x) \, v_{i}.
\end{equation*}

\subsubsection{Directional Derivatives.}

The total derivative of a function $f$ at a point $x$ applied to a vector $\mathbf{v} \in E(X)$,
or more compactly $J(x)(\mathbf{v})$, quantifies how much the output of $f$ varies as $x$ is 
translated infinitesimally in the direction of $v$.  Consequently 
$J(x)(\mathbf{v}) = \mathbf{J}(x) \cdot \mathbf{v}$ is called the \emph{forward directional derivative}.

\begin{figure*}
\centering
\begin{tikzpicture}[scale=0.35, thick]

\draw [rounded corners=2pt, color=black] (-5, 0) rectangle +(-14, 10);
  \node[] at (-12, -1) { $X$ };

  \draw [rounded corners=2pt, color=black] (5, 0) rectangle +(14, 10);
  \node[] at (12, -1) { $Y$ };
  
  \node at (0, -1) { $f: X \rightarrow Y$ };
  
  \draw [->, >=stealth, color=light, line width=1.5] (-10, 6) -- (10.86, 3.01);

  \draw[->, >=stealth, mid, line width=1.5] (11, 3) -- (14, 7);
  \node[mid] at (16, 4.5) { $\mathbf{J}_{f} \cdot (\mathbf{x}' - \mathbf{x})$ };
  
  \fill [fill=dark, text=dark] (11, 3) circle (0.15)
  node[below] { $f(x)$ };

  \fill [fill=dark, text=dark] (14, 7) circle (0.15)
  node[above] { $\tilde{f}(x')$ };
  
  \draw[->, >=stealth, mid, line width=1.5] (-10, 6) -- (-9, 8);
  \node[mid] at (-12, 7.25) { $\mathbf{x}' - \mathbf{x}$ };
  
  \fill [fill=dark, text=dark] (-10, 6) circle (0.15)
  node[below] { $x$ };
  
  \fill [fill=dark, text=dark] (-9, 8) circle (0.15)
  node[above] { $x'$ };
  
\end{tikzpicture}
\caption{
The forward directional derivative of the function $f$ at $x \in X$, $\mathbf{J}_{f}$,
propagates infinitesimal perturbations to the input, $\mathbf{x}' - \mathbf{x}$, to
infinitesimal perturbations of the output, $\mathbf{J}_{f} \cdot (\mathbf{x}' - \mathbf{x})$.
Translating the function output $f(x)$ by $\mathbf{J}_{f} \cdot (\mathbf{x}' - \mathbf{x})$
generates the best linear approximation to $f$ at $x$, $\tilde{f}$.
}
\label{fig:linear_approximation} 
\end{figure*}
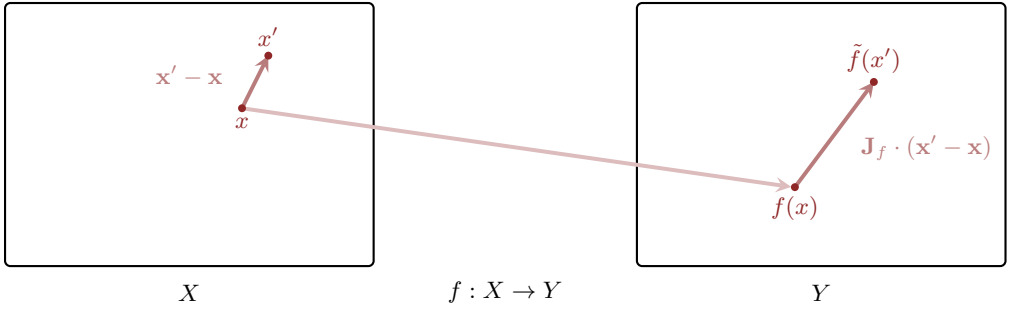

The forward directional derivative is often used to construct the linear function between 
$X$ and $Y$ that best approximates $f$ at $x$.  The best approximation evaluated at $x'$ 
is given by translating $f(x)$ along $J(x) (\mathbf{x}' - \mathbf{x} )$ 
(Figure \ref{fig:linear_approximation})
\begin{align*}
\tilde{f}(x') 
&= f(x) + J(x) (\mathbf{x}' - \mathbf{x} )
\\
&= f(x) + \mathbf{J}(x) \cdot (\mathbf{x}' - \mathbf{x}),
\end{align*}
or in components,
\begin{equation*}
\tilde{f}_{j}(x'_{1}, \ldots, x'_{I}) = f_{j}(x_{1}, \ldots, x_{I})
+ \sum_{i = 1}^{I} \frac{\partial f_{j}}{\partial x_{i}} (x_{1}, \ldots, x_{I})
\cdot (x'_{i} - x_{i}).
\end{equation*}

The total derivative also defines an \emph{adjoint transformation} that maps 
vectors in $E(Y)$ to vectors in $E(X)$, $J^{\dagger}(x) : E(Y) \rightarrow E(X)$.  
The matrix components of this adjoint transformation are given by the transpose of 
the Jacobian matrix,
\begin{equation*}
J^{\dagger}_{ji}(x) = J_{ij}(x).
\end{equation*}
The application of this adjoint transformation to a vector 
$\boldsymbol{\alpha} \in E(Y)$,
\begin{equation*}
\boldsymbol{\beta} = J^{\dagger}(x)(\boldsymbol{\alpha}) = \mathbf{J}^{T} \cdot \boldsymbol{\alpha},
\end{equation*}
quantifies how the inputs need to vary around $x$ in order to achieve the given 
output variation $\boldsymbol{\alpha}$.  We will refer to this action as a 
\emph{reverse directional derivative}.

\subsubsection{The Chain Rule.}

The chain rule provides an explicit construction for the total derivative of
a composite function constructed from many component functions.  Consider for 
example a sequence of functions $f_{n} : X_{n} \rightarrow X_{n + 1}$ and the 
composite function 
\begin{equation*}
f = f_{N} \circ \ldots \circ f_{n} \circ \ldots \circ f_{1} :  X_{1} \rightarrow X_{N + 1}.
\end{equation*}

The total derivative of $f$ at any point $x_{1} \in X_{1}$ is given by composing 
the total derivatives of each component function together in the same order,
\begin{equation*}
J_{f} = J_{f_{N}}(x_{N}) \circ \ldots \circ J_{f_{n}}(x_{n}) \circ \ldots \circ J_{f_{1}}(x_{1}),
\end{equation*}
where $x_{n} = f_{n - 1}(x_{n - 1})$.  Likewise the components of the composite 
Jacobian matrix are given by a sequence of matrix products,
\begin{equation*}
\mathbf{J}_{f} = 
\mathbf{J}_{f_{N}}(x_{N}) \cdot \ldots \cdot 
\mathbf{J}_{f_{n}}(x_{n}) \cdot \ldots \cdot
\mathbf{J}_{f_{1}}(x_{1}).
\end{equation*}
Unfortunately these intermediate matrix products are expensive to evaluate, especially
when the intermediate spaces $X_{n}$ are high-dimensional.  Constructing the full 
composite Jacobian matrix is usually computationally burdensome.

On the other hand this composite structure is well-suited to the evaluation of directional 
derivatives.  For example the forward directional derivative of a composite function is 
given by
\begin{equation*}
J_{f}(\mathbf{v}) = 
\big( J_{f_{N}}(x_{N}) \circ \ldots \circ J_{f_{n}}(x_{n}) \circ \ldots \circ J_{f_{1}}(x_{1}) \big)
(\mathbf{v}),
\end{equation*}
or
\begin{equation*}
\mathbf{J}_{f} \cdot \mathbf{v} = 
\big( \mathbf{J}_{f_{N}}(x_{N}) \cdot \ldots \cdot 
\mathbf{J}_{f_{n}}(x_{n}) \cdot \ldots \cdot 
\mathbf{J}_{f_{1}}(x_{1}) \big)
\cdot \mathbf{v}.
\end{equation*}
Because of the associativity of matrix multiplication we can apply each component derivative 
to $\mathbf{v}$ in sequence and avoid the intermediate compositions entirely,
\begin{equation*}
J_{f}(\mathbf{v}) = 
J_{f_{N}}(x_{N}) ( \cdots J_{f_{n}}(x_{n})( \cdots (J_{f_{1}}(x_{1}) (\mathbf{v}) ) \cdots ) \cdots ),
\end{equation*}
or
\begin{equation*}
\mathbf{J}_{f} \cdot \mathbf{v} = 
\mathbf{J}_{f_{N}}(x_{N}) \cdot ( \cdots \mathbf{J}_{f_{n}}(x_{n}) \cdot ( \cdots \mathbf{J}_{f_{1}}(x_{1})
\cdot \mathbf{v} ) \cdots ) \cdots ).
\end{equation*}
In other words we can evaluate the total forward directional derivative iteratively,
\begin{align*}
\mathbf{v}_{1} &= J_{f_{1}}(x_{1})(\mathbf{v})
\\
\mathbf{v}_{2} &= J_{f_{2}}(x_{2})(\mathbf{v}_{1})
\\
\ldots
\\
\mathbf{v}_{n} &= J_{f_{n}}(x_{n})(\mathbf{v}_{n - 1})
\\
\ldots
\\
\mathbf{v}_{N - 1} &= J_{f_{N - 1}}(x_{N - 1})(\mathbf{v}_{N - 2})
\\
J_{f}(x)(\mathbf{v}) = \mathbf{v}_{N} &= J_{f_{N}}(x_{N})(\mathbf{v}_{N - 1}),
\end{align*}
or in components,
\begin{align*}
\mathbf{v}_{1} &= \mathbf{J}_{f_{1}}(x_{1}) \cdot \mathbf{v}
\\
\mathbf{v}_{2} &= \mathbf{J}_{f_{2}}(x_{2}) \cdot \mathbf{v}_{1}
\\
\ldots
\\
\mathbf{v}_{n} &= \mathbf{J}_{f_{n}}(x_{n}) \cdot \mathbf{v}_{n - 1}
\\
\ldots
\\
\mathbf{v}_{N - 1} &= \mathbf{J}_{f_{N - 1}}(x_{N - 1}) \cdot \mathbf{v}_{N - 2}
\\
\mathbf{J}_{f}(x) \cdot \mathbf{v} = \mathbf{v}_{N} &= \mathbf{J}_{f_{N}}(x_{N}) \cdot \mathbf{v}_{N - 1}.
\end{align*}
The evaluation of each intermediate directional derivative requires only 
a matrix-vector product which is substantially less expensive to implement 
than the matrix-matrix products needed to evaluate the composite Jacobian.

The reverse directional derivative can be evaluated sequentially as well,
\begin{align*}
\boldsymbol{\alpha}_{N} &= J^{\dagger}_{f_{N}}(x_{N})(\boldsymbol{\alpha})
\\
\boldsymbol{\alpha}_{N - 1} &= J^{\dagger}_{f_{N - 1}}(x_{N - 1})(\boldsymbol{\alpha}_{N})
\\
\ldots
\\
\boldsymbol{\alpha}_{N - n} &= J^{\dagger}_{f_{N - n}}(x_{N - n})(\boldsymbol{\alpha}_{N - n + 1})
\\
\ldots
\\
\boldsymbol{\alpha}_{2} &= J^{\dagger}_{f_{2}}(x_{2})(\boldsymbol{\alpha}_{3})
\\
J^{\dagger}_{f}(x)(\boldsymbol{\alpha}) = \boldsymbol{\alpha}_{1} &= J^{\dagger}_{f_{1}}(x_{1})(\boldsymbol{\alpha}_{2}),
\end{align*}
or in components,
\begin{align*}
\boldsymbol{\alpha}_{N} &= \mathbf{J}^{T}_{f_{N}}(x_{N}) \cdot \boldsymbol{\alpha}
\\
\boldsymbol{\alpha}_{N - 1} &= \mathbf{J}^{T}_{f_{N - 1}}(x_{N - 1}) \cdot \boldsymbol{\alpha}_{N}
\\
\ldots
\\
\boldsymbol{\alpha}_{N - n} &= \mathbf{J}^{T}_{f_{N - n}}(x_{N - n}) \cdot \boldsymbol{\alpha}_{N - n + 1}
\\
\ldots
\\
\boldsymbol{\alpha}_{2} &= \mathbf{J}^{T}_{f_{2}}(x_{2}) \cdot \boldsymbol{\alpha}_{3}
\\
\mathbf{J}^{T}_{f}(x) \cdot \boldsymbol{\alpha} = \boldsymbol{\alpha}_{1} &= \mathbf{J}^{T}_{f_{1}}(x_{1}) \cdot \boldsymbol{\alpha}_{2}.
\end{align*}

\subsection{Express Yourself}

Before we can consider how to implement derivatives of real-valued functions in practice 
we first have to consider how to implement the real-valued functions themselves.  Functions 
$f: X \rightarrow Y$ are often implemented as computer programs which take in any input 
value $x \in X$ and return the corresponding output value $f(x) \in Y$.  These computer 
programs are themselves implemented as a sequence of \emph{expressions}, each of which 
transforms some descendent of the initial value towards the final output value.  

If these expressions defined full component functions then we could use the chain rule 
to automatically propagate directional derivatives through the program as we discussed 
in the previous section.  Well-defined component functions, however, would depend on 
only the output of the previous component function, while expressions can depend on the 
output of multiple previous expressions.  Because of this expressions do not immediately 
define valid component functions on their own, and the chain rule does not immediately 
apply.

\begin{figure*}
\centering
\begin{tikzpicture}[scale=0.2, thick]

  \pgfmathsetmacro{\r}{2.5}

  \begin{scope}[shift={(0, 0)}]
    \draw[white] (-16, -4) rectangle (16, 24);
    
    \node at (0, 22) { Pseudo-Code };
  
    \node[anchor=west] at (-15, 8) { \algo };

  \end{scope}

  \begin{scope}[shift={(32, 0)}]
    \draw[white] (-16, -4) rectangle (16, 24);
    
    \node at (0, 22) { Expression Graph };
  
    \coordinate (A) at (-8, 0);
    \coordinate (B) at (0, 0);
    \coordinate (C) at (8, 0);

    \coordinate (D) at (-4, 8);
    \coordinate (E) at (4, 8);
    
    \coordinate (F) at (0, 16);

    \foreach \B/\E in {A/D, B/D, B/E, C/E, D/F, E/F} {
      \draw[-{Stealth[length=6pt, width=6pt]}, shorten <=12.1, shorten >=15, 
            color=mid, line width=1.5] (\B) -- (\E);
    }

    \filldraw[fill=white, draw=dark, line width=1.5] (A) circle (\r)
    node[color=black] { $x_{1, 1}$ };

    \filldraw[fill=white, draw=dark, line width=1.5] (B) circle (\r)
    node[color=black] { $x_{1, 2}$ };
    
    \filldraw[fill=white, draw=dark, line width=1.5] (C) circle (\r)
    node[color=black] { $x_{1, 3}$ };
  
    \filldraw[fill=dark, draw=dark, line width=1.5] (D) circle (\r)
    node[color=white] { $+$ };
    
    \filldraw[fill=dark, draw=dark, line width=1.5] (E) circle (\r)
    node[color=white] { $*$ };
    
    \filldraw[fill=dark, draw=dark, line width=1.5] (F) circle (\r)
    node[color=white] { $/$ };

  \end{scope}
  
\end{tikzpicture}
\caption{
This pseudo-code implements a function $f : \mathbb{R}^{3} \rightarrow \mathbb{R}$ 
that maps every input $x_{1} = (x_{1, 1}, x_{1, 2}, x_{1, 3})$ to a real-valued  
output through three intermediate expressions.  The dependencies between these 
expressions and the input variables forms an expression graph.
}
\label{fig:expr_graph} 
\end{figure*}
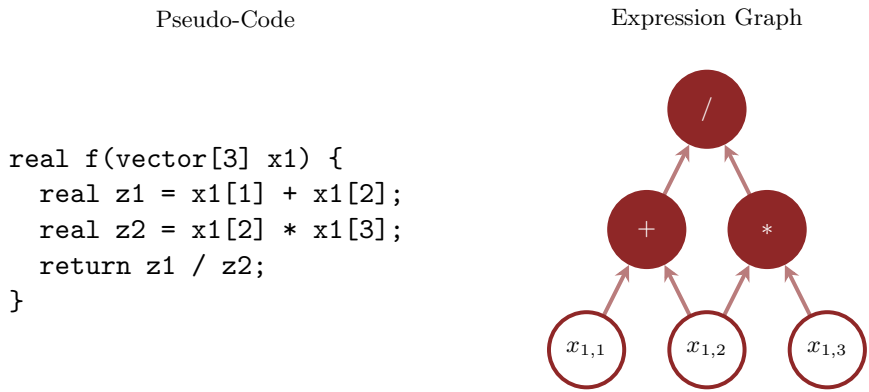

That said we can manipulate each expression into a well-defined component function 
with a little bit of work.  First we'll need to take advantage of the fact that the 
expressions that comprise a complete program can be represented as a directed acyclic 
graph, also known as an \textit{expression graph} (Figure \ref{fig:expr_graph}). In
each expression graph the root nodes designate the input variables, internal and leaf 
nodes designate the expressions, and edges designate the dependencies between the 
expressions.

\begin{myverbbox}[]{\algo}
real f(vector[3] x1) {
  real z1 = x1[1] + x1[2];
  real z2 = x1[2] * x1[3];
  return z1 / z2;
}
\end{myverbbox}

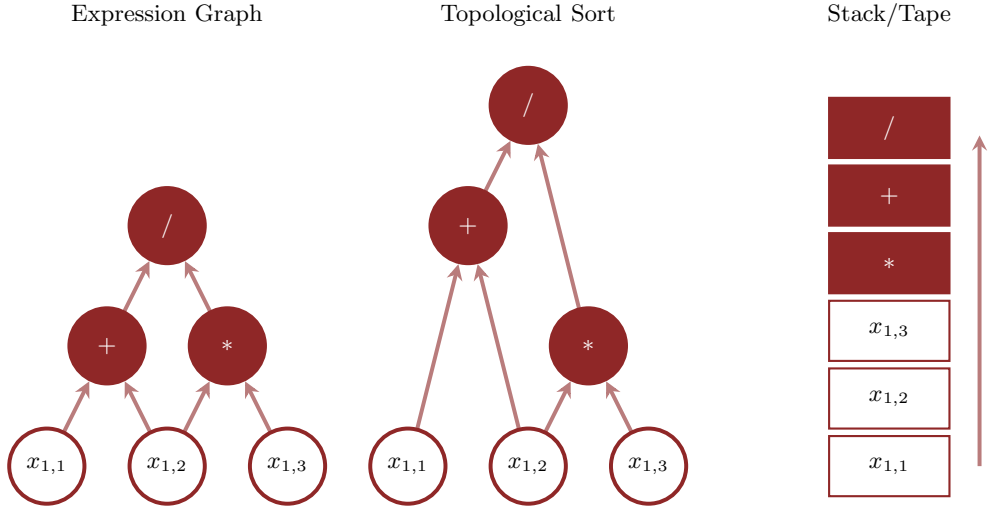
\begin{figure*}
\centering
\begin{tikzpicture}[scale=0.2, thick]

  \pgfmathsetmacro{\r}{2.5}

  \begin{scope}[shift={(0, 0)}]
    \draw[white] (-12, -4) rectangle (12, 32);
    
    \node at (0, 30) { Expression Graph };
  
    \coordinate (A) at (-8, 0);
    \coordinate (B) at (0, 0);
    \coordinate (C) at (8, 0);

    \coordinate (D) at (-4, 8);
    \coordinate (E) at (4, 8);
    
    \coordinate (F) at (0, 16);

    \foreach \B/\E in {A/D, B/D, B/E, C/E, D/F, E/F} {
      \draw[-{Stealth[length=6pt, width=6pt]}, shorten <=12.1, shorten >=15, 
            color=mid, line width=1.5] (\B) -- (\E);
    }

    \filldraw[fill=white, draw=dark, line width=1.5] (A) circle (\r)
    node[color=black] { $x_{1, 1}$ };

    \filldraw[fill=white, draw=dark, line width=1.5] (B) circle (\r)
    node[color=black] { $x_{1, 2}$ };
    
    \filldraw[fill=white, draw=dark, line width=1.5] (C) circle (\r)
    node[color=black] { $x_{1, 3}$ };
  
    \filldraw[fill=dark, draw=dark, line width=1.5] (D) circle (\r)
    node[color=white] { $+$ };
    
    \filldraw[fill=dark, draw=dark, line width=1.5] (E) circle (\r)
    node[color=white] { $*$ };
    
    \filldraw[fill=dark, draw=dark, line width=1.5] (F) circle (\r)
    node[color=white] { $/$ };

  \end{scope}

  \begin{scope}[shift={(24, 0)}]
    \draw[white] (-12, -4) rectangle (12, 32);
    
    \node at (0, 30) { Topological Sort };
  
    \coordinate (A) at (-8, 0);
    \coordinate (B) at (0, 0);
    \coordinate (C) at (8, 0);

    \coordinate (D) at (-4, 16);
    \coordinate (E) at (4, 8);
    
    \coordinate (F) at (0, 24);

    \foreach \B/\E in {A/D, B/D, B/E, C/E, D/F, E/F} {
      \draw[-{Stealth[length=6pt, width=6pt]}, shorten <=12.1, shorten >=15, 
            color=mid, line width=1.5] (\B) -- (\E);
    }

    \filldraw[fill=white, draw=dark, line width=1.5] (A) circle (\r)
    node[color=black] { $x_{1, 1}$ };

    \filldraw[fill=white, draw=dark, line width=1.5] (B) circle (\r)
    node[color=black] { $x_{1, 2}$ };
    
    \filldraw[fill=white, draw=dark, line width=1.5] (C) circle (\r)
    node[color=black] { $x_{1, 3}$ };
  
    \filldraw[fill=dark, draw=dark, line width=1.5] (D) circle (\r)
    node[color=white] { $+$ };
    
    \filldraw[fill=dark, draw=dark, line width=1.5] (E) circle (\r)
    node[color=white] { $*$ };
    
    \filldraw[fill=dark, draw=dark, line width=1.5] (F) circle (\r)
    node[color=white] { $/$ };

  \end{scope}
  
  \begin{scope}[shift={(48, 0)}]
    \draw[white] (-12, -4) rectangle (12, 32);
    
    \node at (0, 30) { Stack/Tape };
 
    \filldraw[draw=dark, fill=white] (-4, -2) rectangle +(8, 4)
    node[midway] { $x_{1, 1}$ };
    
    \filldraw[draw=dark, fill=white] (-4, +2.5) rectangle +(8, 4)
    node[midway] { $x_{1, 2}$ };

    \filldraw[draw=dark, fill=white] (-4, +7) rectangle +(8, 4)
    node[midway] { $x_{1, 3}$ };
    
    \filldraw[draw=dark, fill=dark] (-4, +11.5) rectangle +(8, 4)
    node[midway, white] { $*$ };
    
     \filldraw[draw=dark, fill=dark] (-4, +16) rectangle +(8, 4)
    node[midway, white] { $+$ };
    
    \filldraw[draw=dark, fill=dark] (-4, +20.5) rectangle +(8, 4)
    node[midway, white] { $/$ };
  
    \draw[->, >=stealth, color=mid, line width=1.5] (6, 0) -- (6, 22); 
    
  \end{scope}

\end{tikzpicture}
\caption{
A topological sort of an expression graph is an ordering of the expressions, 
often called a \emph{stack} or a \emph{tape}, that guarantees that when 
progressing across the stack in order each expression will not be evaluated 
until all of the expressions on which it depends have already been evaluated.
}
\label{fig:topological_sort} 
\end{figure*}

A \emph{topological sort} of an expression graph is any ordering of the nodes such 
that each expression follows all expressions on which it depends 
(Figure \ref{fig:topological_sort}).  Such a sorting ensures that if we process
the sorted expressions in order then we will evaluate an expression only once all 
of the expressions on which it depends have already been evaluated.

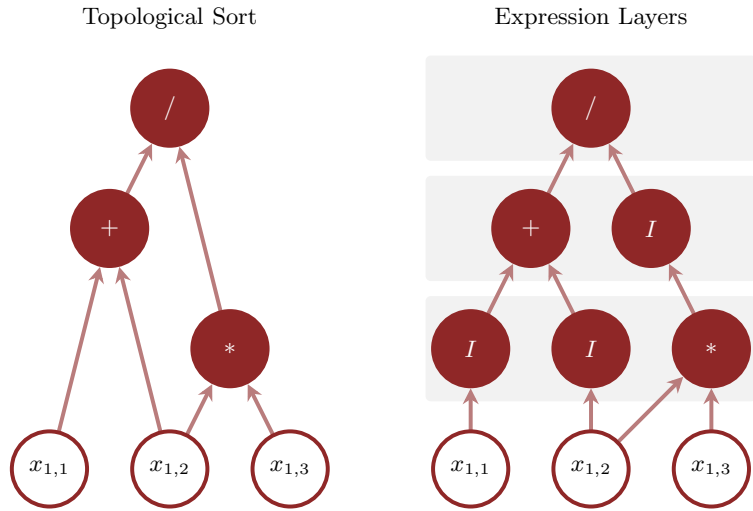
\begin{figure*}
\centering
\begin{tikzpicture}[scale=0.2, thick]

  \pgfmathsetmacro{\r}{2.5}

  \begin{scope}[shift={(0, 0)}]
    \draw[white] (-12, -4) rectangle (12, 32);
    
    \node at (0, 30) { Topological Sort };
  
    \coordinate (A) at (-8, 0);
    \coordinate (B) at (0, 0);
    \coordinate (C) at (8, 0);

    \coordinate (D) at (-4, 16);
    \coordinate (E) at (4, 8);
    
    \coordinate (F) at (0, 24);

    \foreach \B/\E in {A/D, B/D, B/E, C/E, D/F, E/F} {
      \draw[-{Stealth[length=6pt, width=6pt]}, shorten <=12.1, shorten >=15, 
            color=mid, line width=1.5] (\B) -- (\E);
    }

    \filldraw[fill=white, draw=dark, line width=1.5] (A) circle (\r)
    node[color=black] { $x_{1, 1}$ };

    \filldraw[fill=white, draw=dark, line width=1.5] (B) circle (\r)
    node[color=black] { $x_{1, 2}$ };
    
    \filldraw[fill=white, draw=dark, line width=1.5] (C) circle (\r)
    node[color=black] { $x_{1, 3}$ };
  
    \filldraw[fill=dark, draw=dark, line width=1.5] (D) circle (\r)
    node[color=white] { $+$ };
    
    \filldraw[fill=dark, draw=dark, line width=1.5] (E) circle (\r)
    node[color=white] { $*$ };
    
    \filldraw[fill=dark, draw=dark, line width=1.5] (F) circle (\r)
    node[color=white] { $/$ };

  \end{scope}

  \begin{scope}[shift={(28, 0)}]
    \draw[white] (-12, -4) rectangle (12, 32);
    
    \node at (0, 30) { Expression Layers };
  
    \fill[rounded corners=2pt, color=gray95] (-11, 8 - 3.5) rectangle +(22, 7);
    \fill[rounded corners=2pt, color=gray95] (-11, 16 - 3.5) rectangle +(22, 7);
    \fill[rounded corners=2pt, color=gray95] (-11, 24 - 3.5) rectangle +(22, 7);
  
    \coordinate (A) at (-8, 0);
    \coordinate (B) at (0, 0);
    \coordinate (C) at (8, 0);

    \coordinate (D) at (-8, 8);
    \coordinate (E) at (0, 8);
    \coordinate (F) at (8, 8);
  
    \coordinate (G) at (-4, 16);
    \coordinate (H) at (4, 16);
    
    \coordinate (I) at (0, 24);

    \foreach \B/\E in {A/D, B/E, B/F, C/F, D/G, E/G, F/H, G/I, H/I} {
      \draw[-{Stealth[length=6pt, width=6pt]}, shorten <=12.1, shorten >=15, 
            color=mid, line width=1.5] (\B) -- (\E);
    }

    \filldraw[fill=white, draw=dark, line width=1.5] (A) circle (\r)
    node[color=black] { $x_{1, 1}$ };

    \filldraw[fill=white, draw=dark, line width=1.5] (B) circle (\r)
    node[color=black] { $x_{1, 2}$ };
    
    \filldraw[fill=white, draw=dark, line width=1.5] (C) circle (\r)
    node[color=black] { $x_{1, 3}$ };
  
    \filldraw[fill=dark, draw=dark, line width=1.5] (D) circle (\r)
    node[color=white] { $I$ };
    
    \filldraw[fill=dark, draw=dark, line width=1.5] (E) circle (\r)
    node[color=white] { $I$ };
    
    \filldraw[fill=dark, draw=dark, line width=1.5] (F) circle (\r)
    node[color=white] { $*$ };

    \filldraw[fill=dark, draw=dark, line width=1.5] (G) circle (\r)
    node[color=white] { $+$ };
    
    \filldraw[fill=dark, draw=dark, line width=1.5] (H) circle (\r)
    node[color=white] { $I$ };
    
    \filldraw[fill=dark, draw=dark, line width=1.5] (I) circle (\r)
    node[color=white] { $/$ };

  \end{scope}

\end{tikzpicture}
\caption{
In order to turn each topologically-sorted expression into a valid component 
function they must be complemented with identity maps that carry forward 
intermediate values needed by future expressions.  The resulting layers of
expressions depend on only expressions in the previous layer.
}
\label{fig:encapsulated_expressions} 
\end{figure*}

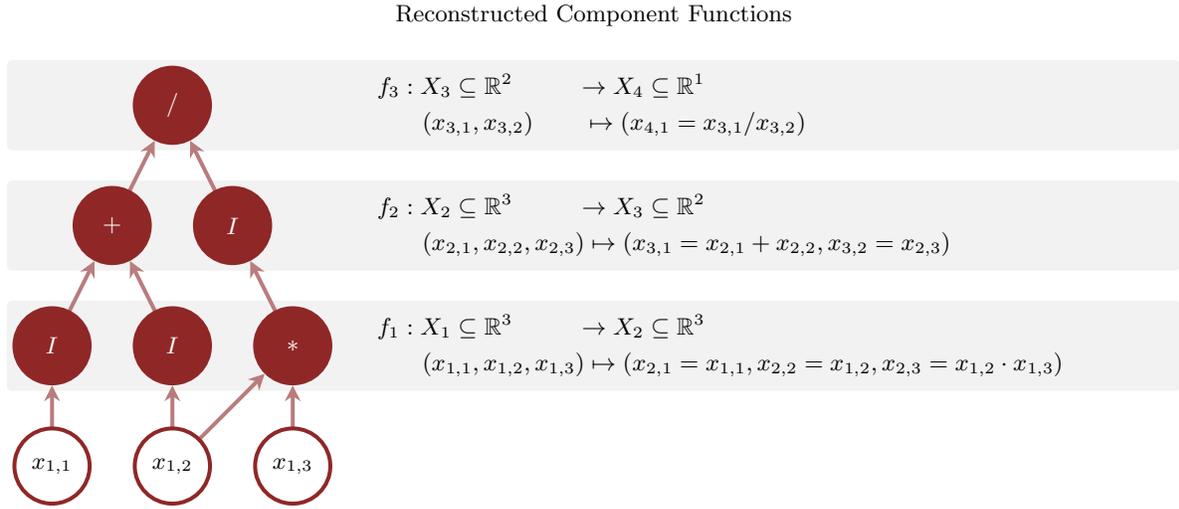
\begin{figure*}
\centering
\begin{tikzpicture}[scale=0.2, thick]

  \pgfmathsetmacro{\r}{2.5}

  \begin{scope}[shift={(0, 0)}]
    \draw[white] (-12, -4) rectangle (68, 32);
    
    \node at (28, 30) { Reconstructed Component Functions };
  
    \fill[rounded corners=2pt, color=gray95] (-11, 8 - 3) rectangle +(78, 6);
    \fill[rounded corners=2pt, color=gray95] (-11, 16 - 3) rectangle +(78, 6);
    \fill[rounded corners=2pt, color=gray95] (-11, 24 - 3) rectangle +(78, 6);

    \node[anchor=west] at (13, 9.25) 
      { $f_{1} : X_{1} \subseteq \mathbb{R}^{3} \quad\quad\;\; \rightarrow X_{2} \subseteq \mathbb{R}^{3}$ };
    
    \node[anchor=west] at (16, 6.75) 
      { $(x_{1, 1}, x_{1, 2}, x_{1, 3}) \mapsto 
         (x_{2, 1} = x_{1, 1}, x_{2, 2} = x_{1, 2}, x_{2, 3} = x_{1, 2} \cdot x_{1, 3})$ };

    \node[anchor=west] at (13, 17.25) 
      { $f_{2} : X_{2} \subseteq \mathbb{R}^{3} \quad\quad\;\; \rightarrow X_{3} \subseteq \mathbb{R}^{2}$ };
    
    \node[anchor=west] at (16, 14.75) 
      { $(x_{2, 1}, x_{2, 2}, x_{2, 3}) \mapsto 
         (x_{3, 1} = x_{2, 1} + x_{2, 2}, x_{3, 2} = x_{2, 3})$ };
    
    \node[anchor=west] at (13, 25.25) 
      { $f_{3} : X_{3} \subseteq \mathbb{R}^{2} \quad\quad\;\; \rightarrow X_{4} \subseteq \mathbb{R}^{1}$ };
                 
    \node[anchor=west] at (16, 22.75) 
      { $(x_{3, 1}, x_{3, 2}) \quad\quad \mapsto (x_{4, 1} = x_{3, 1} / x_{3, 2})$ };
                     
    \coordinate (A) at (-8, 0);
    \coordinate (B) at (0, 0);
    \coordinate (C) at (8, 0);

    \coordinate (D) at (-8, 8);
    \coordinate (E) at (0, 8);
    \coordinate (F) at (8, 8);
  
    \coordinate (G) at (-4, 16);
    \coordinate (H) at (4, 16);
    
    \coordinate (I) at (0, 24);

    \foreach \B/\E in {A/D, B/E, B/F, C/F, D/G, E/G, F/H, G/I, H/I} {
      \draw[-{Stealth[length=6pt, width=6pt]}, shorten <=12.1, shorten >=15, 
            color=mid, line width=1.5] (\B) -- (\E);
    }

    \filldraw[fill=white, draw=dark, line width=1.5] (A) circle (\r)
    node[color=black] { $x_{1, 1}$ };

    \filldraw[fill=white, draw=dark, line width=1.5] (B) circle (\r)
    node[color=black] { $x_{1, 2}$ };
    
    \filldraw[fill=white, draw=dark, line width=1.5] (C) circle (\r)
    node[color=black] { $x_{1, 3}$ };
  
    \filldraw[fill=dark, draw=dark, line width=1.5] (D) circle (\r)
    node[color=white] { $I$ };
    
    \filldraw[fill=dark, draw=dark, line width=1.5] (E) circle (\r)
    node[color=white] { $I$ };
    
    \filldraw[fill=dark, draw=dark, line width=1.5] (F) circle (\r)
    node[color=white] { $*$ };

    \filldraw[fill=dark, draw=dark, line width=1.5] (G) circle (\r)
    node[color=white] { $+$ };
    
    \filldraw[fill=dark, draw=dark, line width=1.5] (H) circle (\r)
    node[color=white] { $I$ };
    
    \filldraw[fill=dark, draw=dark, line width=1.5] (I) circle (\r)
    node[color=white] { $/$ };

  \end{scope}

\end{tikzpicture}
\caption{
Complementing each topologically-sorted expression with the appropriate 
identity maps defines valid component functions.  When composed together
these component functions yield the function implemented by the computer 
program, here $f = f_{3} \circ f_{2} \circ f_{1}$, and allow for the 
application of the chain rule to differentiate through the program.
}
\label{fig:component_functions} 
\end{figure*}

Any topological sort provides an explicit sequence of expressions, but each
expression can still depend on the output of any expression that precedes it
in the stack.  We can use the ordering to limit this dependence to only the 
previous output, however, if we can buffer each expression with any of the 
previous outputs that are used by future expressions.  For example this 
buffering can be implemented by introducing identify expressions that 
propagate any necessary values forward (Figure \ref{fig:encapsulated_expressions}).  
Together each initial expression and the added identify expressions define a 
\emph{layer} of expressions that depend on only the expressions in the previous 
layer, so that these layers define a sequence of valid component functions 
(Figure \ref{fig:component_functions}).

Once we've derived these component functions we can finally apply the chain 
rule.  In particular we can evaluate forward directional derivatives by 
propagating an initial vector through the constructed sequence of component 
functions.  At the same time we can evaluate a reverse directional derivative 
by first evaluating all of the component functions in a forward sweep through 
the sequence before executing a reverse sweep that propagates a vector in the 
output space to the input space.

Something interesting happens, however, when we evaluate the total derivative 
of one of these reconstructed component functions.  Let's denote the action 
of the component function as $f_{n} : (x, x') \mapsto (g(x), I(x'))$, where 
$g$ is the action implemented by the initial expression and $I$ is the 
identify map that propagates any auxiliary outputs.  We can also decompose
the input vector $\mathbf{v}$ into components that map onto $g$ and $I$,
respectively,
\begin{equation*}
\mathbf{v} = (\mathbf{v}_{g}, \mathbf{v}_{I})^{T}.
\end{equation*}
In this notation the forward directional derivative becomes
\begin{align*}
J_{f_{n}} (\mathbf{v}) 
&=
J_{f_{n}} ( (\mathbf{v}_{g}, \mathbf{v}_{I})^{T} )
\\
&=
\mathbf{J}_{f_{n}} \cdot (\mathbf{v}_{g}, \mathbf{v}_{I})^{T}
\\
&=
\renewcommand\arraystretch{1.5}
\begin{pmatrix}
\frac{\partial g}{\partial x} & \frac{\partial g}{\partial x'} 
\\
\frac{\partial I}{\partial x} & \frac{\partial I}{\partial x'}
\end{pmatrix}
\cdot
\begin{pmatrix}
\mathbf{v}_{g} 
\\
\mathbf{v}_{I}
\end{pmatrix}
\\
&=
\renewcommand\arraystretch{1.5}
\begin{pmatrix}
\frac{\partial g}{\partial x} & 0
\\
0 & 0
\end{pmatrix}
\cdot
\begin{pmatrix}
\mathbf{v}_{g}
\\
\mathbf{v}_{I}
\end{pmatrix}
\\
&=
\renewcommand\arraystretch{1.5}
\begin{pmatrix}
\frac{\partial g}{\partial x} \cdot \mathbf{v}_{g}
\\
0
\end{pmatrix}
\end{align*}

The only non-vanishing contribution to the directional derivative comes from 
the initial expression; the vector corresponding to the buffered outputs, 
$\mathbf{v}_{I}$, completely decouples from any derivatives that follow.  
Consequently the only aspect of the reconstructed component function that 
influences propagation of the forward directional derivative is the forward 
directional derivative of the expression itself.  

In other words we can evaluate the forward directional derivative of a function 
implemented by a computer program by propagating only the forward directional 
derivatives of the individual expressions; at no point do we actually have to 
construct explicit component functions!  The action of the adjoint derivative 
behaves similarly, allowing us to evaluate the reverse directional derivative 
using only the reverse directional derivatives local to each expression.  When 
considering higher-order derivatives, however, the equivalence between 
propagating vectors across expressions and full component functions is not 
always preserved and explicit component functions may be needed.  For more 
see \cite{Betancourt:2018b}.

\subsection{Automatic Differentiation}

Automatic differentiation exploits the equivalence between propagating 
vectors across expressions and full component functions to algorithmically 
evaluate forward and reverse directional derivatives.

\emph{Forward mode automatic differentiation} implements forward 
directional derivatives, passing intermediate values and intermediate
forward directional derivatives between expressions as the program is
evaluated.  These intermediate forward directional derivatives are 
denoted \emph{tangents} or \emph{sensitivities}.

Any implementation of a function $g$ that supports forward mode automatic 
differentiation must provide not only a map from input values to output
values but also a map from input tangents to output tangents,
\begin{equation*}
\mathbf{v}' = J_{g} \cdot \mathbf{v}.
\end{equation*}

Similarly \emph{reverse mode automatic differentiation} implements
reverse directional derivatives.  Here the program must be evaluated
first before intermediate reverse directional derivatives are 
propagated between expressions in the \emph{reverse} order that 
the expressions are evaluated.  These intermediate reverse directional
derivatives are denoted \emph{cotangents} or \emph{adjoints}.

Any implementation of a function $g$ that supports reverse mode automatic 
differentiation must provide not only a map from input values to output 
values but also a map from output cotangents to input cotangents,
\begin{equation*}
\boldsymbol{\alpha}' = J^{T}_{g} \cdot \boldsymbol{\alpha}.
\end{equation*}

\section{Automatic Differentiation of Finite-Dimensional Implicit Functions} \label{sec:finite}
A finite collection of constraint functions that an output has to satisfy for 
a given input implicitly defines a map from inputs to outputs, or an implicit 
function.  In this section we discuss how implicit functions are formally 
defined, how to differentiate these implicitly defined functions, and then 
finally demonstrate their application on several instructive examples.

\subsection{The Finite-Dimensional Implicit Function Theorem}

Consider a finite-dimensional real-valued space of known inputs, $X$, a 
finite-dimensional real-valued space of unknown outputs, $Y$, a finite-dimensional 
real-valued space of constraint values, $Z$, and the \textit{constraint function}
\begin{alignat*}{6}
c :\; &X \times Y & &\rightarrow& \; &Z&
\\
&(x, y)& &\mapsto& &c(x, y)&.
\end{alignat*}
The implicit function theorem defines the conditions under which the constraint 
function implicitly defines a map from $f : X \to Y$ which satisfies $c(x, f(x)) = 0$ 
for all inputs in a local neighborhood $x \in U \subset X$.

More formally consider the neighborhoods $0 \in W \subset Z$ and $V \subset Y$ such 
that the kernel of the constraint function falls into product of $U$ and $V$,
\begin{equation*}
c^{-1}(0) = U \times V,
\end{equation*}
and assume that a function $f : U \rightarrow V$ that satisfies $c(x, f(x)) = 0$ 
exists.  If the constraint function $c$ is differentiable across $U \times V$ then 
the total derivative of $c$ evaluated at $(x, f(x))$ is given by 
\begin{align*}
0 
&= \frac{\text d}{\text d x} c(x, f(x)) 
\\
&= \frac{\partial c}{\partial x} (x, f(x)) 
+ \frac{\partial c}{\partial y} (x, f(x)) \circ \frac{\mathrm{d} f}{\mathrm{d} x} (x).
\end{align*}

When the dimension of $Z$ equals the dimension of $Y$ then the partial derivative
$\partial c / \partial y \ (x, y)$ might define a bijection from $V$ to $W$.  If it does then
 we can solve for the total derivative of the assumed function,
\begin{equation*}
\underbrace{\frac{ \text{d} f }{ \text{d} x } (x)}_{U \rightarrow V}
=
- \underbrace{\left( \frac{ \partial c }{ \partial y } (x, f(x)) \right)^{-1}}_{W \rightarrow V}
\circ 
\underbrace{ \frac{ \partial c }{ \partial x } (x, f(x))}_{U \rightarrow W}.
\end{equation*}
When $\partial c / \partial y \ (x, y)$ is bijective this derivative is well-defined, and the existence and uniqueness of ordinary
differential equation solutions guarantees that an implicit function that satisfies 
$c(x, f(x)) = 0$ is well-defined in the neighborhood around $x$.  In other words 
the system of constraints defines an implicit function if and only if the partial 
derivative $\partial c / \partial y \ (x, y)$ is invertible

The implicit function theorem determines when an implicit function is well-defined,
but not how to evaluate it.  In practice we typically have to rely on numerical 
methods that heuristically search the output space for a value $y$ that satisfies
$c(x, y) = 0$ for the given input $x$.  

\subsection{Evaluating Directional Derivatives of Finite-Dimensional Implicit Functions}

To incorporate implicit functions into an automatic differentiation library we 
need to be able to evaluate not only the output consistent with a given input 
but also the directional derivatives.  Here we consider three general approaches: 
a \textit{trace method} that works with a given numerical solver, a method that 
utilizes intermediate results of the implicit function theorem, and an 
adjoint method that evaluates the directional derivative directly.

In all three approaches we will consider not the implicit function alone but rather 
its composition with a summary function that maps the outputs into some real space, 
$g : Y \rightarrow \mathbb{R}^{K}$.  Often $g$ will be the identify map, but the
flexibility offered by this summary function will facilitate some of the examples 
that we consider below.  

When $X = \mathbb{R}^{I}$ and $Y = Z = \mathbb{R}^{J}$ the composition of the summary
function with the implicit function defines the real-valued function
\begin{equation*}
h = g \circ f : \mathbb{R}^{I} \rightarrow \mathbb{R}^{J} \rightarrow \mathbb{R}^{K}
\end{equation*}
and our goal will be to evaluate either the forward directional derivative 
$J_{g \circ f}(x)(\mathbf{v})$ or the reverse directional derivative 
$J^{\dagger}_{g \circ f}(x)(\boldsymbol{\alpha})$.

\subsubsection{Trace Method.} \label{sec:trace}

Each step that an iterative numerical solver takes while searching for a consistent 
output can be interpreted as a map from the output space to itself given the fixed 
input $x$,
\begin{alignat*}{6}
\tilde{f}_{n} :\; &X \times Y& &\rightarrow& \; &Y&
\\
&(x, y_{n - 1})& &\mapsto& &y_{n}&.
\end{alignat*}
The \emph{trace} of the solver's evaluation then defines a composite function,
\begin{equation*}
\tilde{f}(x, y_{0}) = 
\big( \tilde{f}_{N}(x) \circ \tilde{f}_{N - 1}(x) \circ \ldots \circ f_{1}(x) \big) (y_{0})
\end{equation*}
that maps the input and an initial guess to an approximate solution,
\begin{alignat*}{6}
\tilde{f} :\; &X \times Y& &\rightarrow& \; &Y&
\\
&(x, y_{0})& &\mapsto& &\tilde{y}&
\end{alignat*}
satisfying $c(x, \tilde{y}) \approx 0$.

If each of these intermediate steps are differentiable and supported by an automatic
differentiation library then we can evaluate the directional derivatives of
$\tilde{f}$ using automatic differentiation and use them to approximate the directional 
derivatives of the exact implicit function $f$.  While straightforward to implement 
this approach can suffer from poor performance in practice, especially as the 
number of solver iterations grows and propagating derivatives through the composite 
function becomes slow and memory intensive.  See for example \citep{Bell:2008} for 
further discussion of this trace method applied to optimization problems and 
\citep{Margossian:2019} for one on algebraic equations.

\subsubsection{Finite-Dimensional Implicit Function Theorem.} \label{sec:finite_ift}

Conveniently the derivative of the implicit function $f$ is explicitly constructed
in the derivation of the implicit function theorem,
\begin{equation*}
\frac{ \text{d} f }{ \text{d} x } (x)
=
- \left( \frac{ \partial c }{ \partial y } (x, f(x)) \right)^{-1}
\circ 
\frac{ \partial c }{ \partial x } (x, f(x)).
\end{equation*}
If we can evaluate these derivatives of the constraint function then we can immediately
evaluate the derivative for $f$ once we have numerically solved for the output of the implicit 
function $y = f(x)$,
\begin{equation*}
\frac{ \text{d} f }{ \text{d} x } (x)
=
- \left( \frac{ \partial c }{ \partial y } (x, y) \right)^{-1}
\circ 
\frac{ \partial c }{ \partial x } (x, y).
\end{equation*}

The Jacobian of the composite function $g \circ f$ is then given by
\begin{align*}
J_{g \circ f}(x)
&=
\frac{ \partial (g \circ f) }{ \partial x }(x)
\\
&=
\frac{ \mathrm{d} g }{ \mathrm{d} y} (f(x)) \circ \frac{ \mathrm{d} f }{ \mathrm{d} x} (x)
\\
&= -
\frac{ \mathrm{d} g }{ \text{d} y} (f(x))
\circ 
\left( \frac{ \partial c }{ \partial y } (x, f(x)) \right)^{-1}
\circ 
\frac{ \partial c }{ \partial x_{i} } (x, f(x)).
\end{align*}

In components this becomes
\begin{align*}
(J_{g \circ f})_{ik}(x)
&=
\frac{ \partial h_{k} }{ \partial x_{i} }(x)
\\
&=
\sum_{j = 1}^{J}
\frac{ \mathrm{d} g_{k} }{ \mathrm{d} y_{j}} (f(x)) \cdot 
\frac{ \mathrm{d} f_{j} }{ \mathrm{d} x_{i}} (x)
\\
&= -
\sum_{j = 1}^{J}
\sum_{j' = 1}^{J}
\frac{ \text{d} g_{k} }{ \text{d} y_{j}} (f(x))
\circ 
\left( \frac{ \partial c_{j'} }{ \partial y_{j} } (x, f(x)) \right)^{-1}
\circ 
\frac{ \partial c_{j'} }{ \partial x_{i} } (x, f(x)),
\end{align*}
or in more compact matrix notation,
\begin{equation*}
\mathbf{J}_{g \circ f} = - \mathbf{J}_g \cdot \mathbf{C}_y^{-1} \cdot \mathbf C_x,
\end{equation*} 
where
\begin{align*}
(\mathbf{C}_{y})_{ij} &= \frac{\partial c_i}{\partial y_j} (x, f(x))
\\
(\mathbf{C}_{x})_{ij} &= \frac{\partial c_i}{\partial x_j} (x, f(x))
\\
(\mathbf{J}_{g})_{ij} &= \frac{\partial g_j}{\partial y_i} (f(x)).
\end{align*}

In order to incorporate this composite function into forward mode automatic 
differentiation we need to evaluate the action of the Jacobian contracted
against a tangent vector,
\begin{align*}
\left (\mathbf{J}_{g \circ f}(x) \cdot \mathbf v \right)_j
&=
\sum_{i = 1}^{I} (J_{g \circ f})_{ij}(x) \cdot v_{i}
\\
&=
- \sum_{i = 1}^{I} 
\frac{ \text{d} g_{j} }{ \text{d} y} (f(x))
\circ 
\left( \frac{ \partial c }{ \partial y } (f(x)) \right)^{-1}
\circ 
 \frac{ \partial c }{ \partial x_i } (x)
\cdot v_i,
\end{align*}
or equivalently
\begin{equation*}
\mathbf J_{g \circ f}(x) \cdot \mathbf v 
= - \mathbf J_g \cdot \mathbf C_y^{-1} \cdot \mathbf C_x \cdot \mathbf v.
\end{equation*}
Similarly to incorporate this composite function into reverse mode automatic
differentiation we need to evaluate the action of the Jacobian contracted
against a cotangent vector,
\begin{align*}
(\mathbf J_{g \circ f}^T(x) \cdot \boldsymbol{\alpha})_i
&=
\sum_{j = 1}^{J} (J_{g \circ f})_{ij}(x) \cdot \alpha_{j}
\\
&=
- \sum_{j = 1}^{J}
\frac{ \text d g_{j} }{ \text d y} (f(x))
\circ 
\left( \frac{ \partial c }{ \partial y } (f(x)) \right)^{-1}
\circ 
\frac{ \partial c }{ \partial x } (f(x)) \cdot \alpha_{j},
\end{align*}
or
\begin{eqnarray*}
\mathbf J_{g \circ f}^T(x) \cdot \boldsymbol \alpha 
& = & - \mathbf C_x^T \cdot \left (\mathbf C_y^{-1} \right)^T \cdot \mathbf J_g^T \cdot \boldsymbol \alpha \\
& = & - \mathbf C_x^T \cdot \left (\mathbf C_y^T \right)^{-1} \cdot \mathbf J_g^T \cdot \boldsymbol \alpha.
\end{eqnarray*}

With so many terms there are multiple ways to evaluate these directional 
derivatives.  The \textit{forward} method, for example, explicitly constructs 
$\mathbf{J}_{g \circ f} (x) = - \mathbf J_g \cdot \mathbf C_y^{-1} \cdot \mathbf C_x$
before evaluating the contractions $\mathbf{J}_{g \circ f} (x) \cdot \mathbf{v}$ or 
$\mathbf{J}_{g \circ f}^T(x) \cdot \boldsymbol{\alpha}$.  With careful use of automatic 
differentiation, however, we can evaluate the final directional derivatives 
more efficiently.  In particular we don't need to explicitly construct $\mathbf J_g$ 
and $\mathbf C_x$ at all.

For example when evaluating $\mathbf{J}_{g \circ f} (x) \cdot \mathbf{v}$ we can avoid
$\mathbf C_x$ by using one sweep of forward mode automatic differentiation to 
evaluate $\mathbf{u} = \mathbf C_x \cdot \mathbf{v}$ directly.  Because of the 
inversion we have to construct the entirety of $\mathbf{C}_{y}$, for 
example with $J$ sweeps of forward mode or reverse mode automatic 
differentiation, but we can avoid constructing $(\mathbf{C}_{y})^{-1}$
by solving for only the linear system
\begin{equation*}
\mathbf{C}_{y} \cdot \mathbf{t} = \mathbf{u}.
\end{equation*}
Finally we can avoid constructing $\mathbf{J}_{g}$ with one sweep of forward 
mode automatic differentiation to evaluate $\mathbf{J}_{g} \cdot \mathbf{t}$.
These steps are outlined in Algorithm~\ref{alg:forward}.

\begin{algorithm}
  \caption{Forward mode automatic differentiation of finite-dimensional implicit function.}
  \label{alg:forward}
  \begin{algorithmic}[1]
    \item \textbf{input:} constraint function, $c$; summary function, $g$; tangent, $\mathbf v$.
    \textit{Note: we assume we have a differentiable program for all the input functions.}
    \item $\mathbf u = \mathbf C_x \cdot \mathbf v$ (one forward mode sweep)
    \item $\mathbf C_y = \partial c / \partial y$ ($J$ forward or reverse mode sweep)
    \item $\mathbf t  = \mathbf C_y^{-1} \cdot \mathbf u$ (linear solve)
    \item $\mathbf J_{g \circ f} \cdot \mathbf v = - \mathbf J_g \cdot \mathbf t$ (one forward sweep)
    \item \textbf{return:} $\mathbf J_{g \circ f} \cdot \mathbf v$
  \end{algorithmic}
\end{algorithm}

Similar optimizations are also possible when evaluating the reverse directional 
derivative $\mathbf J_{g \circ f}^T (x) \cdot \boldsymbol \alpha$.  We first evaluate
$\boldsymbol{\beta} = \mathbf{J}_{g}^{T} \cdot \boldsymbol{\alpha}$ using
one sweep of reverse mode automatic differentiation and then construct
$\mathbf{C}_{y}$ as above.  This allows us to solve the linear system
\begin{equation*}
\mathbf{C}^{T}_{y} \cdot \boldsymbol{\gamma} = \boldsymbol{\beta},
\end{equation*}
and then evaluate $\mathbf{C}^T_x \cdot \boldsymbol{\beta}$ with one more
sweep of reverse mode automatic differentiation through the constraint 
function.
These steps are outlined in Algorithm~\ref{alg:reverse}.

\begin{algorithm}
  \caption{Reverse mode automatic differentiation of finite-dimensional implicit function.}
  \label{alg:reverse}
  \begin{algorithmic}[1]
    \item \textbf{input:} constraint function, $c$; summary function, $g$; cotangent, $\boldsymbol \alpha$.
    \textit{Note: we assume we have a differentiable program for all the input functions.}
    \item $\boldsymbol \beta = \mathbf J^T_g \cdot \boldsymbol \alpha$ (one reverse mode sweep)
    \item $\mathbf C_y = \partial c / \partial y$ ($J$ forward or reverse mode sweep)
    \item $\boldsymbol \gamma = \left ( \mathbf C_y^T \right)^{-1} \boldsymbol \beta$ (linear solve)
    \item $\mathbf J^T_{g \circ f} \cdot \boldsymbol \alpha = - \mathbf C^T_x \cdot \boldsymbol \gamma$ (one reverse mode sweep)
    \item \textbf{return:} $\mathbf J^T_{g \circ f} \cdot \boldsymbol \alpha$
  \end{algorithmic}
\end{algorithm}

Often the form of a particular constraint function results in sparsity 
structure in $\mathbf{C}_{x}$ and $\mathbf{C}_{y}$ that can be exploited 
to reduce the cost of the irreducible linear algebraic operations.
Identifying this structure and implementing faster operations, however,
requires substantial experience with numerical linear algebra.

\subsubsection{The Finite-Dimensional Adjoint Method.} \label{sec:adjoint}

The adjoint method provides a way of directly implementing the contractions
that give forward and reverse directional derivatives without explicitly 
constructing the Jacobian $ \text{d} f / \text{d} x (x)$.  Adjoint methods 
have historically been constructed for specific implicit functions but
here we present a general construction for any finite-dimensional system
of constraints.

We begin by constructing a binary \emph{Lagrangian} function,
\begin{equation*}
\mathcal{L} : X \times Y \rightarrow \mathbb{R},
\end{equation*}
whose derivative is equal to the desired contraction when $c(x, y) = 0$.  
More formally we compose $\mathcal{L}$ with the implicit function defined 
by the constraints to give the unary function
\begin{equation*}
\mathcal{L}(x) = \mathcal{L}(x, f(x)) : x \mapsto \mathbb{R},
\end{equation*}
and then require that 
\begin{equation*}
\frac{ \mathrm{d} \mathcal{L} }{ \mathrm{d} x_{i} }(x)
=
(\mathbf{J}_{g \circ f} \cdot \mathbf v)_{i} (x).
\end{equation*}
for forward mode automatic differentiation or 
\begin{equation*}
\frac{ \mathrm{d} \mathcal{L} }{ \mathrm{d} x_{i} }(x)
=
( \mathbf J^T_{g \circ f} \cdot \boldsymbol \alpha )_{i} (x)
\end{equation*}
for reverse mode automatic differentiation.  To simplify the presentation 
from here on we will focus on only this latter application to reverse mode 
automatic differentiation.

Next we introduce \emph{any} mapping $\Lambda: Z \rightarrow \mathbb{R}$ that
preserves the kernel of the constraint function, $\Lambda \circ c(x, y) = 0$ 
whenever $c(x, y) = 0$.  This allows us to define a second function
\begin{equation*}
\mathcal{Q} = \Lambda \circ c : X \times Y \rightarrow \mathbb{R}
\end{equation*}
with $\mathcal{Q}(x, f(x)) = 0$ for all $x \in X$.

These two functions together then define an \emph{augmented Lagrangian} function,
\begin{equation*}
\mathcal{J} = \mathcal{L} + \mathcal{Q} : X \times Y \rightarrow \mathbb{R}.
\end{equation*}

Substituting $y = f(x)$ gives a unary function,
\begin{align*}
\mathcal{J}(x) 
&= \mathcal{J}(x, f(x)) 
\\
&= \mathcal{L}(x, f(x)) + \mathcal{Q}(x, f(x)).
\end{align*}
The second term, however, vanishes by construction so that $\mathcal{J}(x)$
reduces to $\mathcal{L}(x)$ and 
\begin{equation*}
\frac{ \mathrm{d} \mathcal{J} }{ \mathrm{d} x_{i} }(x)
=
\frac{ \mathrm{d} \mathcal{L} }{ \mathrm{d} x_{i} }(x)
=
(\mathbf{J}^T_{g \circ f} \cdot \boldsymbol \alpha)_{i} (x).
\end{equation*}

While the contribution from $\mathcal{C}(x, f(x))$ vanishes, its inclusion
into the augmented Lagrangian introduces another way to evaluate the desired
directional derivative.  The total derivative of the unary augmented Lagrangian 
is
\begin{align*}
\frac{ \mathrm{d} \mathcal{J} }{ \mathrm{d} x_{i} }(x)
&=
\frac{ \mathrm{d} \mathcal{J} }{ \mathrm{d} x_{i} }(x, f(x))
\\
&=
\frac{ \mathrm{d} \mathcal{L} }{ \mathrm{d} x_{i} }(x, f(x))
+
\frac{ \mathrm{d} \mathcal{Q} }{ \mathrm{d} x_{i} }(x, f(x))
\\
&= \quad
\frac{ \partial \mathcal{L} }{ \partial x_{i} }(x, f(x))
+ \left( \frac { \partial \mathcal{L} }{ \partial y } (x, f(x)) \circ \frac{ \partial f}{ \partial x_{i} } \right)(x)
\\
& \quad
+ \frac{ \partial \mathcal{Q} }{ \partial x_{i} }(x, f(x))
+ \left( \frac { \partial \mathcal{Q} }{ \partial y } (x, f(x)) \circ \frac{ \partial f}{ \partial x_{i} } \right)(x)
\\
&= \quad
\frac{ \partial \mathcal{L} }{ \partial x_{i} }(x, f(x))
+ \frac{ \partial \mathcal{Q} }{ \partial x_{i} }(x, f(x))
\\
& \quad
+ \left( \left( \frac { \partial \mathcal{L} }{ \partial y } (x, f(x))
+ \frac { \partial \mathcal{Q} }{ \partial y } (x, f(x)) \right) \circ \frac{ \partial f}{ \partial x_{i} } \right)(x).
\end{align*}
Once we've identified the solution $y = f(x)$ the first two terms are ordinary 
partial derivatives that are straightforward  to evaluate. The second two 
terms, however, are tainted by the total derivative of the implicit function,
$\mathrm{d} f / \mathrm{d} x$, which is expensive to evaluate as we saw in the
previous section.

At this point, however, we can exploit the freedom in the choice of kernel-preserving 
function $\Lambda$ and hence $\mathcal{J}$ itself.  If we can engineer a $\Lambda$
such that
\begin{equation*}
 \frac { \partial \mathcal{L} }{ \partial y } (x, f(x)) 
+ \frac { \partial \mathcal{Q} }{ \partial y } (x, f(x)) = 0
\end{equation*}
then the contribution from the last two terms, and the explicit dependence 
on the derivative of the implicit function vanishes entirely!

The adjoint method attempts to solve this \emph{adjoint system}
\begin{equation*}
\frac { \partial \mathcal{L} }{ \partial y } (x, f(x))
+ \frac { \partial \Lambda }{ \partial c } (c(x, f(x)))
\circ \frac { \partial c }{ \partial y } (x, f(x))
= 0
\end{equation*}
for a suitable $\Lambda$ and then evaluate the desired contraction from the
remaining two terms,
\begin{equation*}
\frac{ \mathrm{d} \mathcal{J} }{ \mathrm{d} x_{i} }(x)
=
\frac{ \partial \mathcal{L} }{ \partial x_{i} }(x, f(x))
+ \frac { \partial \Lambda }{ \partial c } (c(x, f(x)))
\circ \frac { \partial c }{ \partial x_{i} } (x, f(x)).
\end{equation*}
If a suitable $\Lambda$ exists, and we can find it, then this two-stage method allows us to 
evaluate the directional derivative directly without constructing the 
derivatives of the implicit function.

\subsection{Demonstrations}

To compare and contrast the two presented methods for evaluating directional 
derivatives of an implicit function we examine their application on three
common implicit systems: algebraic equations, difference equations, and 
optimization problems.

\subsubsection{Algebraic systems.} \label{sec:algebraic}

When $X = \mathbb{R}^{I}$ and $Y = \mathbb{R}^{J}$ a system of $J$ transverse constraint 
functions $c_{j}(x, y)$ defines an algebraic system and a well-defined implicit function.  
In Section \ref{sec:finite_ift} we saw that the finite-dimensional implicit function 
theorem gives the reverse directional derivative
\begin{equation*}
\mathbf J^T(x) \cdot \boldsymbol \alpha 
= - \mathbf C_x^T \cdot \left (\mathbf C_y^{-1} \right)^T \cdot \mathbf J_g^T \cdot \boldsymbol \alpha.
\end{equation*}
where
\begin{align*}
(\mathbf{J}_{g})_{ij} &= \frac{\partial g_j}{\partial y_i} (f(x))
\\
(\mathbf{C}_{y})_{ij} &= \frac{\partial c_i}{\partial y_j} (x, f(x))
\\
(\mathbf{C}_{x})_{ij} &= \frac{\partial c_i}{\partial x_j} (x, f(x)).
\end{align*}
Here we will take $g$ to be the identify function so that $\mathbf{J}_{g}$ reduces to 
the identify matrix and the reverse directional derivative simplifies to
\begin{equation*}
\mathbf J^T(x) \cdot \boldsymbol \alpha 
= - \mathbf C_x^T \cdot \left (\mathbf C_y^{-1} \right)^T \cdot \boldsymbol \alpha.
\end{equation*}

To apply the adjoint method we need to construct a Lagrangian function that satisfies
\begin{equation*}
\frac{ \mathrm{d} \mathcal{L} }{ \mathrm{d} x_{i} }(x, f(x))
=
( \mathbf{J}^T \cdot \boldsymbol \alpha)_{i} (x).
\end{equation*}
For example we can take
\begin{equation*}
\mathcal{L} 
= \mathbf f^T(x) \cdot \boldsymbol \alpha 
= \sum_{j = 1}^{J} f_{j}(x) \cdot \alpha_{j}.
\end{equation*}

Next we need to augment $\mathcal{L}$ with the contribution from the constraints $\mathcal{Q}$.  
Because inner products vanish whenever either input is zero we can take
\begin{equation*}
\mathcal{Q} 
= \mathbf c^T(x, y) \cdot \boldsymbol \lambda 
= \sum_{j = 1}^{J} c_{j}(x, y) \cdot \lambda_{j},
\end{equation*}
for any real-valued, non-zero constants $\lambda_{j}$.

With these choices of $\mathcal{L}$ and $\mathcal{Q}$ the adjoint system becomes
\begin{align*}
0
&=
\frac{ \partial \mathcal{L} }{ \partial y_{j} }
+ \frac{ \partial \mathcal{Q} }{ \partial y_{j} }
\\
&=
\alpha_{j} 
+ \sum_{j' = 1}^{J} \frac{ \partial c_{j'} }{ \partial y_{j} }(x, f(x)) \cdot \lambda_{j'},
\end{align*}
or, in matrix notation,
\begin{equation*}
0 = \boldsymbol{\alpha} + \mathbf{C}^T_{y} \cdot \boldsymbol{\lambda}.
\end{equation*}
Because $\mathbf{C}_{y}$ is non-singular we can directly solve this for 
$\boldsymbol{\lambda}$,
\begin{equation*}
\boldsymbol \lambda = - (\mathbf{C}_{y}^{-1})^T \cdot \boldsymbol \alpha.
\end{equation*}

Substituting this into the remaining terms then gives
\begin{align*}
(\mathbf{J}^T \cdot \boldsymbol \alpha)_{i} (x)
&=
\frac{ \mathrm{d} \mathcal{J} }{ \mathrm{d} x_{i} }(x, f(x))
\\
&= \frac{ \partial \mathcal{L} }{ \partial x_{i} }(x, f(x))
+ \frac{ \partial \mathcal{Q} }{ \partial x_{i} }(x, f(x))
\\
&= 0
+ \sum_{j' = 1}^{J} \lambda_{j'} \cdot \frac{ \partial c_{j'} }{ \partial x_{i} }(x, f(x))
\\
&= (\mathbf{C}_{x}^{T} \cdot \boldsymbol \lambda)_{i}
\\
&= (- \mathbf{C}_{x}^{T} \cdot (\mathbf{C}_{y}^{-1})^T \cdot \boldsymbol \alpha)_{i},
\end{align*}
or
\begin{equation*}
\mathbf{J}^T \cdot \boldsymbol \alpha
= - \mathbf C_x^T \cdot (\mathbf C_y^{-1})^T \cdot \mathbf J_g^T \cdot \boldsymbol \alpha.
\end{equation*}
which is exactly the same as the result from the implicit function theorem.

In this general case there isn't any structure in the constraint functions to 
exploit and consequently both methods yield equivalent calculations.

\subsubsection{Difference Equations.} \label{sec:de}

To better contrast the two methods let's consider a more structured system.  A
discrete dynamical system over the state space $\mathbb{R}^{N}$ defines trajectories
\begin{equation*}
y = (\mathbf{y}_1, \ldots, \mathbf{y}_{i}, \ldots, \mathbf{y}_{I}),
\end{equation*}
where the individual states $\mathbf{y}_{i} \in \mathbb{R}^{N}$ are implicitly
defined by the difference equations
\begin{equation*}
\mathbf{y}_{i + 1} - \mathbf{y}_i = \mathbf{\Delta}(\mathbf{y}_i, x, i),
\end{equation*}
for some initial condition $\mathbf{y}_0 = \mathbf u(x)$.  To simplify the notation
we will write
\begin{equation*}
\mathbf{\Delta}_i = \mathbf{\Delta}(\mathbf{y}_i, x, i)
\end{equation*}
from here on.

Organizing these difference equations into constraint equations defines a highly 
structured algebraic system,
\begin{align*}
\mathbf{y}_{1} - \mathbf{y}_0 - \mathbf{\Delta}_0 &= c_{1}(x, y) = 0 
\\
\cdots&
\\
\mathbf{y}_{i} - \mathbf{y}_{i - 1} - \mathbf{\Delta}_{i - 1} &= c_{i}(x, y) = 0
\\
\cdots&
\\
\mathbf{y}_{I} - \mathbf{y}_{I - 1} - \mathbf{\Delta}_{I - 1} &= c_{I}(x, y) = 0,
\end{align*}
which then sets the stage for the implicit function machinery.

Formally this system of constraints defines an entire trajectory; the output 
space is given by $Y \subset \mathbb R^{N \times I}$.  Often, however, we are
interested not in the entire trajectory but only the final state, 
$\mathbf{y}_{I} \in \mathbb{R}^{N}$.  Conveniently we can readily accommodate 
this by using the summary function to project out the final state,
\begin{alignat*}{6}
g :\; &Y = \mathbb{R}^{N \times I} & &\rightarrow& \; &\mathbb{R}&
\\
&(\mathbf{y}_1, \ldots, \mathbf{y}_{i}, \ldots, \mathbf{y}_{I})& &\mapsto& &\mathbf{y}_{I}&.
\end{alignat*}

Having defined an implicit system and summary function we can now apply the 
implicit function theorem and adjoint methods to derive the gradients of 
the implicitly-defined final state.  Although these two methods yield 
equivalent results, the adjoint method more directly incorporates the natural 
structure of the problem without any explicit linear algebra.

\noindent
\textit{Differentiation with the Implicit Function Theorem.}
To apply the implicit function theorem directly we proceed as in \ref{sec:algebraic} and
and compute
\begin{equation*}
\mathbf J^T \cdot \boldsymbol \alpha 
= - \mathbf C^T_x \cdot (\mathbf C_y^{-1})^T \cdot \mathbf J_g^T \cdot \boldsymbol \alpha.
\end{equation*}
term by term from the right.

Our chosen summary function yields the Jacobian matrix
\begin{equation*}
\mathbf J_g 
= \frac{\partial g}{\partial \mathbf y} 
= \left [ \mathbf 0_N, \cdots, \mathbf 0_N, \mathbf I_N \right ],
\end{equation*}
where $\mathbf 0_N$ is an $N \times N$ matrix of zeros and $\mathbf I_N$ is the $N \times N$ 
identity matrix.  The first contraction on the right then gives
\begin{equation*}
{\boldsymbol \beta}^T 
= (\mathbf J_g^T \cdot \boldsymbol \alpha)^T
= [ \underbrace{0, 0, \cdots, 0}_{(I - 1)N}, \underbrace{\alpha_1, \alpha_2, \cdots, \alpha_N}_N ].
\end{equation*}

At this point we construct $\mathbf C_y$ from the derivatives of the constraint functions,
\begin{equation*}
\mathbf C_y = 
  \begin{bmatrix}
    \mathbf I_N & \mathbf 0_N & \cdots \\
    \mathbf C_y^1 & \mathbf I_N & \mathbf 0_N & \cdots \\
    \mathbf 0_N & \mathbf C_y^2 & \mathbf I_N & \mathbf 0_N & \cdots \\
    \vdots & \ddots & \ddots & \ddots & \ddots \\
    \mathbf 0_N & \cdots & \cdots & \mathbf C_y^{I - 1} & \mathbf I_N 
  \end{bmatrix},
\end{equation*}
where
\begin{equation*}
  \mathbf C_y^i = \frac{\partial}{\partial \mathbf y_i} (\mathbf y_{i + 1} - \mathbf y_{i} - \mathbf{\Delta}_i )
    = - 1 - \frac{\partial \mathbf{\Delta}_i}{\partial \mathbf y_i},
\end{equation*}
and then solve the linear system
\begin{equation*}
\mathbf C_y^T \cdot \boldsymbol \gamma = \boldsymbol \beta.
\end{equation*}

Because of the structure of the constraints the matrix $\mathbf C_y$ is triangular and
with enough linear algebra proficiency we would know that we can efficiently solve for 
$\boldsymbol{\gamma}$ with backwards elimination.  To clarify the derivation we first 
split the elements of $\boldsymbol \gamma$ into subvectors of length $N$,
\begin{equation*}
\boldsymbol{\gamma}^{T} = [ \boldsymbol \gamma_1, ..., \boldsymbol \gamma_I ].
\end{equation*}
The linear system can then be written as
\begin{equation*}
  \begin{bmatrix}
    \mathbf I_N & \mathbf C_y^1& \mathbf 0_N & \cdots \\
    \mathbf 0_N & \mathbf I_N &  \mathbf C_y^2  & \mathbf 0_N & \cdots \\
    \mathbf 0_N & \mathbf 0_N & \mathbf I_N & \mathbf C_y^3 & \mathbf 0_N \\
    \vdots & \ddots & \ddots & \ddots & \ddots  \\
    \mathbf 0 & \cdots & \cdots &\mathbf 0 & \mathbf I_N 
  \end{bmatrix}
 \cdot
 \begin{bmatrix}
 \boldsymbol \gamma_1 \\
 \boldsymbol \gamma_2 \\
 \boldsymbol \gamma_3 \\ 
 \vdots \\
 \boldsymbol \gamma_I
 \end{bmatrix}
 =
 \begin{bmatrix}
   0 \\
   0 \\
   0 \\
   \vdots \\
   \boldsymbol \alpha
 \end{bmatrix}.
\end{equation*} 
Starting from the bottom and going up we can solve this system recursively to obtain the 
\textit{backward difference equations}
\begin{equation*}
  \boldsymbol \gamma_i - \left ( 1 + \frac{\partial \mathbf{\Delta}_i}{\partial y_i} \right ) \boldsymbol \gamma_{i + 1} = 0
\end{equation*}
with terminal condition
\begin{equation*}
  \boldsymbol \gamma_I = \boldsymbol \alpha.
\end{equation*}

We now evaluate $\mathbf C_x$; for $i > 1$ the derivatives are straightforward,
\begin{equation*}
(\mathbf C_x)_{i} 
=
\frac{\partial c_i}{\partial x} 
= - \frac{\partial \mathbf{\Delta}_{i - 1}}{\partial x},
\end{equation*}
but for $i = 1$ we have to be careful to incorporate the implicit dependence of the
initial state, $\mathbf y_0 = \mathbf u(x)$, on $x$,
\begin{align*}
(\mathbf C_x)_{1} 
&=
\frac{\partial c_1}{\partial x} 
\\
&= 
\frac{\partial}{\partial x} 
\left( \mathbf y_1 - \mathbf y_0 - \mathbf{\Delta}(\mathbf y_0, x, 0) \right)
\\
&=
\frac{\partial}{\partial x} 
\left( \mathbf y_1 - \mathbf u(x) - \mathbf{\Delta}(\mathbf u(x), x, 0) \right)
\\
&=
- \frac{\partial \mathbf u}{\partial x} 
- \left(\frac{\partial \mathbf{\Delta}}{\partial x} 
   + \frac{\partial \mathbf{\Delta}}{\partial \mathbf u} \cdot \frac{\partial \mathbf u}{\partial x} \right)  
\\
&=
- \left (1 + \frac{\partial \mathbf{\Delta}}{\partial \mathbf u} \right) 
\frac{\partial \mathbf u}{\partial x}
- \frac{\partial \mathbf{\Delta}}{\partial x}.
\end{align*}

Finally we multiply $\mathbf C_x$ and $\boldsymbol \gamma$ to give
\begin{align*}
\left (\frac{\text d \mathbf y_I}{\text d x} \right)^T \cdot \boldsymbol \alpha
&= - \mathbf C_x^T \cdot \boldsymbol \gamma
\\
&= 
\left ( \frac{ \partial \mathbf u}{\partial x} \right)^T \cdot 
\left (1 + \frac{\partial \mathbf{\Delta}_0}{\partial \mathbf u} \right)^T \cdot \boldsymbol \gamma_1 
+ 
\sum_{i = 1}^I \left (\frac{\partial \mathbf{\Delta}_{i - 1}}{\partial x} \right)^T 
\cdot \boldsymbol \gamma_i.
\end{align*}

Although the steps are straightforward, implementing them correctly, let alone efficiently,
has required careful organization.

\noindent
\textit{Differentiation with the Adjoint Method.}
Because this discrete dynamical system is a special case of an algebraic system we could
appeal to the augmented Lagrangian that we constructed in Section~\ref{sec:algebraic}
and then repeat the same linear algebra needed for the implicit function theorem method.
A more astute choice of Lagrangian, however, allows us to directly exploit the structure
of the constraints and the summary function.

Given our choice of summary function we need to construct a Lagrangian function which 
satisfies 
\begin{equation*}
\frac{\partial \mathcal L}{\partial x_i} (x, \mathbf y_I) 
= \left (\frac{\mathrm d \mathbf y_I}{\mathrm d x_i} \right)^T \cdot \boldsymbol \alpha.
\end{equation*}
Because $\mathbf{\Delta}_i$ is a telescoping series,
\begin{equation*}
\mathbf y_I = \mathbf u(x) + \sum_{i = 1}^{I} \mathbf{\Delta}_i,
\end{equation*}
a natural choice is
\begin{align*}
\mathcal L(x, \mathbf y) 
&= 
\mathbf y_I^{T} \cdot \boldsymbol{\alpha}
\\
&=
\mathbf u^T(x) \cdot \boldsymbol \alpha 
+ \sum_{i = 1}^{I} \mathbf{\Delta}_i^T \cdot \boldsymbol \alpha.
\end{align*}

For the constraint term $\mathcal C$ we utilize a similar form as in the general algebraic
case,
\begin{align*}
\mathcal Q(x, y)
&=
\sum_{i = 1}^{I} \mathbf{c}_{i}^{T} \cdot \boldsymbol{\lambda}_{i}
\\
&=
\sum_{i = 1}^{I} [\mathbf y_{i} - \mathbf y_{i - 1} - \mathbf{\Delta}_{i - 1}]^T \cdot \boldsymbol \lambda_i,
\end{align*}
where $\boldsymbol{\lambda}_{i} \in \mathbb{R}^{N}$.

The adjoint system defined by $\mathcal{L}$ and $\mathcal{Q}$ decouples into 
the equations
\begin{align*}
0 &=
\frac{\partial \mathcal L}{\partial \mathbf y_i} 
+ \frac{\partial \mathcal Q}{\partial \mathbf y_i}
\\
&= 
\left ( \frac{\partial \mathbf{\Delta}_i}{\partial y_i} \right)^T  \cdot \boldsymbol \alpha 
+ \lambda_{i + 1} - \lambda_i
- \left (\frac{\partial \mathbf{\Delta}_i}{\partial y_i} \right)^T \cdot \lambda_i.
\end{align*}
for $i \in \{1, \ldots, I - 1\}$ along with the terminal condition for $i = I$,
\begin{equation*}
0 = \boldsymbol \lambda_{I}.
\end{equation*}
In other words the adjoint system defines a \emph{backward difference equation}
that we can solve recursively from $\lambda_{I}$ to $\lambda_{I - 1}$ all the 
way to $\lambda_{1}$.

Once we have solved for these adjoint states we can substitute them into the 
remaining terms to give the desired directional derivative,
\begin{eqnarray*}
  \left (\frac{\mathrm d \mathbf y_I}{\mathrm d x} \right)^T \cdot \boldsymbol \alpha
    & = & \frac{\partial \mathcal L}{\partial x} + \frac{\partial \mathcal Q}{\partial x} \\
    & = & \left (\frac{\partial \mathbf u}{\partial x} \right)^T \cdot \left (1 + \frac{\partial \mathbf{\Delta}_0}{\partial \mathbf u} \right)^T \cdot (\boldsymbol \alpha - \boldsymbol \lambda_0)
    + \sum_{i = 1}^{I} \left ( \frac{\partial \mathbf{\Delta}_i}{\partial x} \right)^T \cdot (\boldsymbol \alpha - \boldsymbol \lambda_i).
\end{eqnarray*}

As expected the adjoint method has provided an alternative path to the same 
result we obtained with the implicit function theorem.  Indeed matching the 
two expressions for $(\mathrm d \mathbf y / \mathrm d x)^T \cdot \boldsymbol \alpha$,
suggests taking
\begin{equation*}
  \boldsymbol \gamma_{i} = \boldsymbol \alpha - \boldsymbol \lambda_i,
\end{equation*}
in which case the intermediate difference equations that arise in both 
methods are exactly the same as well!  The advantage of the adjoint method 
is that we did not have to construct $\mathbf J_g$, $\mathbf C_y$, or
$\mathbf{C}_{x}$, let alone manage their sparsity to ensure the most 
efficient computation.

Note also that this procedure yields the same result for difference equations
derived in \citep{Betancourt:2020} only with fewer steps, and hence fewer 
opportunities for mistakes.

\subsubsection{Optimization.} \label{sec:optimization}

Another common class of finite-dimensional, implicit functions are defined 
as solutions to optimization problems.  Given an \emph{objective function}
$F : X \times Y \rightarrow \mathbb{R}$ we can define the output of the 
implicit function as the value which maximizes that objective function 
for a given input,
\begin{equation*}
y = \underset{v}{\text{argmax}} \, F(x, v).
\end{equation*}

In a neighborhood $U \times V \subset X \times Y$ where $F(x,-)$ is convex 
for all $x \in U$, this implicit function is also defined by the differential 
constraint 
\begin{equation*}
c(x, y) = \frac{\partial F}{\partial y} (x, y) = 0,
\end{equation*}
which allows us to apply our machinery to evaluate the derivatives of 
this implicit function.  The convexity constraint is key here; without it
the constraint function will identify only general extrema which can
include not only maxima but also minima and saddle points.

When $X \subset \mathbb{R}^{I}$ and $Y \subset \mathbb{R}^{J}$ the differential 
constraint function reduces to an algebraic system and we can directly apply 
the results of Section~\ref{sec:algebraic}.  In particular without any 
assumptions on the objective function there is no difference in the 
implementation of the implicit function theorem or adjoint methods.

We start by setting the summary function to the identity so that 
$\mathbf{J}_{g} = \mathbb{I}$ and
\begin{equation*}
\boldsymbol{\beta} = \mathbf{J}_{g}^{T} \cdot \boldsymbol{\alpha} = \boldsymbol{\alpha}.
\end{equation*}

Next we compute 
\begin{equation*}
(\mathbf C_y)_{ij} 
= \frac{\partial c_i}{\partial y_j}(x, y) 
= \frac{\partial^2 F}{\partial y_i \partial y_j}(x, y),
\end{equation*}
for example analytically or with higher-order automatic differentiation 
\citep{Griewank:2008, Betancourt:2018b}.  Conveniently when the optimization
is implemented with a higher-order numerical method this Hessian matrix will 
already be available at the final solution.  Once $\mathbf C_y$ has been 
constructed we then solve the linear system
\begin{equation*}
\frac{\partial^2 F}{\partial y_i \partial y_j}(x, y) \cdot \boldsymbol{\gamma} = \boldsymbol{\alpha}.
\end{equation*}

Finally we construct
\begin{equation*}
(\mathbf C_x)_{ij} = \frac{\partial c_i}{\partial x_j} = \frac{\partial^2 F}{\partial y_i \partial x_j},
\end{equation*}
and then evaluate
\begin{equation*}
\mathbf J^T\cdot \boldsymbol \alpha 
= - \mathbf C_x^T \cdot \boldsymbol{\gamma}.
\end{equation*}
This final contraction defines a second-order directional derivative 
of the objective function.  Conveniently it can be evaluated with higher-order automatic
differentiation without having to explicitly construct $\mathbf C_x$.

These optimization problems become more sophisticated with the introduction of an 
additional equality constraint so that the output of the implicit function is now
defined by the condition
\begin{equation*}
y = \underset{s}{\text{argmax}} \ F(x, s) \, \text{such that} \, k(x, y) = 0,
\end{equation*}
for the auxiliary constraint function $k: \mathbb R^I \times \mathbb R^J \to \mathbb R^K$.

In order to define an implicit system that consistently incorporates both of these 
constraints we have to augment the output space.  We first introduce the Lagrange 
multipliers $\mu \in M \subset \mathbb R^K$ and the augmented objective function
\begin{alignat*}{6}
\Phi :\; &X \times Y \times M& &\rightarrow& \; &\mathbb{R}&
\\
&(x, y, \mu)& &\mapsto& &F(x, y) + \mu \cdot k (x, y)&.
\end{alignat*}
A consistent solution to the constrained optimization problem is then given by
\begin{equation*}
(y, \mu) = \underset{s, m}{\text{argmax}} \, \Phi(x, s, m).
\end{equation*}
In other words we can incorporate all of the constraints directly on the augmented 
output space $\zeta = (y, \mu)$ and then project back down to the original output 
space using the summary function $g: (y, \mu) \mapsto y$.

Within a sufficiently convex neighborhood we can also define the constrained 
optimization problem with a constraint function over this augmented output space,
\begin{equation*}
c(x, \zeta) = \frac{\partial \Phi}{\partial \zeta} (x, \zeta) = 0,
\end{equation*}
which allows us to apply our methods for evaluating directional derivatives.

From the projective summary function we first construct
\begin{equation*}
\boldsymbol{\beta} = \mathbf{J}_{g}^{T} \cdot \boldsymbol{\alpha}.
\end{equation*}
Next we differentiate the constraint function on the augmented output space,
\begin{equation*}
(\mathbf C_\zeta)_{ij} = \frac{\partial^2 \Phi}{\partial \zeta_i \partial \zeta_j}(x, \zeta),
\end{equation*}
and then solve the linear system
\begin{equation*}
\frac{\partial^2 \Phi}{\partial \zeta_i \partial \zeta_j}(x, \zeta) 
\cdot \boldsymbol{\gamma} = \boldsymbol{\beta}.
\end{equation*}
Finally we construct
\begin{equation*}
(\mathbf C_x)_{ij} = \frac{\partial^2 \Phi}{\partial \zeta_i \partial x_j},
\end{equation*}
and then evaluate
\begin{equation*}
\mathbf J^T\cdot \boldsymbol \alpha 
= - \mathbf C_x^T \cdot \boldsymbol{\gamma}.
\end{equation*}
As before this final contraction defines a second-order directional derivative
that can be directly evaluated with higher-order automatic differentiation,
although this time on the augmented output space.

Inequality constraints introduce an additional challenge.  While the Karush-Kuhn-Tucker 
conditions define an appropriate system of constraints \citep{Karush:1939, Kuhn:1951}, 
the dimension of the constraint space
varies with the input $x$ as different inequality constraints become active.  Moreover even
when the objective function and the additional constraint functions are all smooth the implicit 
function they define might not be differentiable at every $x$.  We leave a detailed treatment of 
this problem to future work.

\section{Automatic Differentiation of Infinite-Dimensional Implicit Functions} \label{sec:infinite}
With care in how derivatives are defined we can immediately generalize the finite
dimensional methods for evaluating directional derivatives of implicit functions
to infinite dimensional systems, for example systems that implicitly define entire 
trajectories, fields, or even probability distributions.  In this section we review 
the basics of differentiable, infinite dimensional spaces and the generalization of 
the implicit function theorem before generalizing the directional derivative evaluation 
methods and demonstrating them on two instructive examples.

\subsection{The Infinite-Dimensional Implicit Function Theorem}

The machinery of differential calculus over the real numbers generalizes quite naturally 
to a Fr\'{e}chet calculus over Banach vector spaces.  In this section we review the key 
concepts and then use them to construct an infinite-dimensional implicit function theorem.
For a more in depth presentation of these topics see for example \cite{Kesavan:2020}.

\subsubsection{Fr\'{e}chet derivatives.}

The \emph{Fr\'{e}chet derivative} generalizes the concept of a derivative that we
introduced for real spaces to the more general Banach vector spaces, or more 
compactly \emph{Banach spaces}.  Banach spaces include not only finite-dimensional Euclidean 
vector spaces but also infinite-dimensional function spaces.

Consider two Banach spaces, $X$ and $Y$, and a function $f : X \rightarrow Y$
mapping between them.  If $f$ is Fr\'{e}chet differentiable then the Fr\'{e}chet
derivative assigns to each input point $x \in X$ a bounded, linear map
\begin{equation*}
\frac{ \delta f }{ \delta x } : X \rightarrow \mathrm{BL}(X, Y),
\end{equation*}
where $\mathrm{BL}(X, Y)$ is the space of bounded, linear functions from $X$
to $Y$.  In other words the Fr\'{e}chet derivative of $f$ evaluated at $x$
defines a bounded, linear map from $X$ to $Y$,
\begin{equation*}
\frac{ \delta f }{ \delta x } (x) : X \rightarrow Y.
\end{equation*}
Note that unlike the total derivative we introduced on real spaces the Fr\'{e}chet
derivative is defined directly on a vector space and so there is no distinction 
between locations and directions.

If $Z$ is a third Banach space then the composition of $f$ with $g: Y \rightarrow Z$ 
defines a map from $X$ to $Z$,
\begin{equation*}
g \circ f : X \rightarrow Z.
\end{equation*}
The Fr\'{e}chet derivative of this composition,
\begin{equation*}
 \frac{ \delta (g \circ f) }{ \delta x } (x) : X \rightarrow Z,
\end{equation*}
follows a chain rule,
\begin{equation*}
 \frac{ \delta (g \circ f) }{ \delta x } (x)
=
\underbrace{ \frac{ \delta g }{ \delta x } (f(x)) }_{Y \rightarrow Z}
\circ
\underbrace{ \frac{ \delta f }{ \delta x } (x)}_{X \rightarrow Y}.
\end{equation*}

A binary map $h: X \times Y \rightarrow Z$ also admits a partial Fr\'{e}chet derivative.
For example the partial Fr\'{e}chet derivative of $h$ with respect to the first input
defines a map
\begin{equation*}
\frac{ \delta h }{ \delta x } : X \times Y \rightarrow \mathrm{BL}(X, Z).
\end{equation*}
Equivalently the partial Fr\'{e}chet derivative of $h$ \emph{evaluated at $(x, y)$}
is a bounded, linear map from $X$ to $Z$,
\begin{equation*}
 \frac{ \delta h }{ \delta x } (x, y) : X \rightarrow Z.
\end{equation*}
Similarly we can define a partial derivative with respect to the second output as
\begin{equation*}
\frac{ \delta h }{ \delta y } : X \times Y \rightarrow \mathrm{BL}(Y, Z)
\end{equation*}
with
\begin{equation*}
 \frac{ \delta h }{ \delta y } (x, y) : Y \rightarrow Z.
\end{equation*}

When a binary map like $h$ is composed with unary maps in each argument then 
we can use the chain rule to define a notion of a total Fr\'{e}chet derivative.  
Let $h : X \times Y \rightarrow Z$, $f : W \rightarrow X$ and $g : W \rightarrow Y$.
The component-wise composition $h(f( \, ), g( \, ) )$ then 
defines a unary map
\begin{equation*}
q = h(f( \, ), g( \, ) ) : W \rightarrow Z,
\end{equation*}
with the corresponding Fr\'{e}chet derivative
\begin{equation*}
 \frac{ \delta q }{ \delta w } (w) : W \rightarrow Z,
\end{equation*}
which decomposes into contributions from each argument,
\begin{equation*}
\frac{ \delta q }{ \delta x } (w) = 
\frac{ \delta h }{ \delta x } (f(w), g(w)) \circ \frac{ \delta f }{ \delta w } (w)
+ \frac{ \delta h }{ \delta y } (f(w), g(w)) \circ \frac{ \delta g }{ \delta w }  (w).
\end{equation*}

\subsubsection{The Implicit Function Theorem.}

The finite dimensional implicit function theorem immediately generalizes to
Banach spaces with the use of Fr\'echet derivatives.  Consider a Banach space 
of known inputs, $X$, a Banach space of unknown outputs, $Y$, a Banach space of 
constraint values, $Z$, and the constraint function
\begin{alignat*}{6}
c :\; &X \times Y & &\rightarrow& \; &Z&
\\
&(x, y)& &\mapsto& &c(x, y)&.
\end{alignat*}
Once again, we examine a particular input, $x \in X$,
and the neighborhoods $x \in U \subset X$, $V \subset Y$ and $W \subset Z$.

If $c$ is Fr\'{e}chet differentiable across $U \times V$ 
then it defines two partial Fr\'{e}chet derivatives
\begin{equation*}
 \frac{ \delta c }{ \delta x } (x, y) : U \rightarrow W
\end{equation*}
and
\begin{equation*}
 \frac{ \delta c }{ \delta y } (x, y) : V \rightarrow W.
\end{equation*}

When $ \delta c / \delta y (y)$ defines a bijection from $V$ to $W$
we can also define a corresponding inverse operator,
\begin{equation*}
\left( \frac{ \delta c }{ \delta y } (x, y) \right)^{-1} : W \rightarrow V.
\end{equation*}
In this case the implicit function theorem guarantees that the kernel of the 
constraint function, $c^{-1}(0)$, implicitly defines a Fr\'{e}chet differentiable 
function from the input space to the output space, $f : U \rightarrow V$ that 
satisfies $c(x, f(x)) = 0$.  

As in the finite-dimensional case we can calculate the Fr\'echet derivative of
this implicit function by differentiating the constraint function,
\begin{eqnarray*}
0 & = & \frac{\delta c}{\delta x} (x) \\
& = & \frac{\delta c}{\delta x} (x, y) + \frac{\delta c}{\delta y} (x, y) \circ \frac{\delta y}{\delta x} (x).
\end{eqnarray*}
Because $\delta c / \delta y$ is invertible at $(x, f(x))$ we can immediately 
solve for $\delta y / \delta x$ to give
\begin{equation*}
\underbrace{ \frac{ \delta f }{ \delta x } (x)}_{U \rightarrow V}
= -
\underbrace{\left( \frac{ \delta c }{ \delta y } (x, f(x)) \right)^{-1}}_{W \rightarrow V}
\circ 
\underbrace{\frac{ \delta c }{ \delta x } (x, f(x))}_{U \rightarrow W}.
\end{equation*}
Moving forward we will denote $J_{f} = \delta f / \delta x (x)$ the Fr\'echet Jacobian.

\subsection{Evaluating Directional Derivatives of Infinite-Dimensional Implicit Functions} \label{sec:infinite}

While we can handle infinite dimensional spaces mathematically any practical automatic 
differentiation implementation will be restricted to finite dimensional inputs and
final outputs.  To that end we will assume that the input space is real, $X = \mathbb R^I$,
while allowing the output space $Y$ to be infinite dimensional, for example corresponding
to a smooth trajectory or a latent field.  We will use the summary function, however, to
project that potentially-infinite dimensional output to a finite-dimensional real space,
$g : Y \rightarrow \mathbb{R}^{J}$.  For example we might consider a dynamical system that 
defines an entire trajectory but project out only the finite-dimensional final state.

In order to implement such a system in a reverse mode automatic differentiation library 
we then need to be able to evaluate the reverse directional derivative
\begin{equation*}
J_{g \circ f}^{\dagger}(x)(\boldsymbol{\alpha}) 
= \big( J^{\dagger}_{f}(x) \circ J^{\dagger}_{g}(f(x)) \big) (\boldsymbol{\alpha}).
\end{equation*}
Because $g \circ f$ is a real-to-real map the total action is given by a matrix-vector product,
\begin{equation*}
J_{g \circ f}^{\dagger}(x)(\boldsymbol{\alpha}) = \mathbf{J}_{g \circ f}^{T} \cdot \boldsymbol{\alpha},
\end{equation*}
but the component operators $J^{\dagger}_{f}(x)$ and $J^{\dagger}_{g}(f(x))$ will not be
unless $Y$ is a finite-dimensional real space.

Here we will consider the generalizations of the three methods for evaluating this
directional derivative that we constructed in Section~\ref{sec:finite}.

\subsubsection{Trace Method.}

Although we can't practically construct a numerical method for an implicit function $f : \mathbb{R}^{I} \rightarrow Y$ 
with infinite-dimensional output space, we often can construct numerical methods for the 
finite-dimensional composition $g \circ f : \mathbb{R}^{I} \rightarrow \mathbb{R}^{J}$.
Numerical integrators for ordinary and partial differential equations, for example,
discretize $Y$ in order to approximate the finite dimensional outputs of $g \circ f$
without having to confront an infinite dimensional space directly.

Because the intermediate calculations of the numerical solver will be finite-dimensional 
we immediately apply the trace method discussed in Section~\ref{sec:trace}, automatically 
differentiating through each iteration of the solve, to approximate 
$J_{g \circ f}(x)(\boldsymbol{\alpha})$.  As in the finite-dimensional case, however, this 
direct approach is often too computationally expensive and memory intensive to be practical.

\subsubsection{Infinite-Dimensional Implicit Function Theorem.}

In theory the reverse directional derivative is given immediately by the implicit 
function theorem,
\begin{align*}
J_{g \circ f}^{\dagger}(x)(\boldsymbol{\alpha}) 
&= 
\big( J^{\dagger}_{f}(x) \circ J^{\dagger}_{g}(f(x)) \big) (\boldsymbol{\alpha})
\\
&=
\left( -
\left( \frac{ \delta c }{ \delta x } (x, f(x)) \right)^{\dagger}
\circ
\left( \left( \frac{ \delta c }{ \delta y } (x, f(x)) \right)^{\dagger}\right)^{-1}
\circ J^{\dagger}_{g}(f(x)) \right) (\boldsymbol{\alpha}).
\end{align*}
Unfortunately whenever $Y$ is infinite-dimensional each of these Fr\'echet derivatives
will be infinite dimensional operators that are difficult, if not impossible, to 
implement in practice.  For example $J^{\dagger}_{g}(f(x))$ maps the finite covector
$\boldsymbol{\alpha}$ to an infinite dimensional space that can't be represented in 
finite memory.

\subsubsection{The Infinite-Dimensional Adjoint Method.} \label{sec:adjoint-inf}

With a careful use of Fr\'echet derivatives the adjoint method defined in Section~\ref{sec:adjoint} 
generalizes to the case where $Y$ and $Z$ are general Banach spaces.  As usual we consider 
a particular input, $x \in X$, and the neighborhoods $x \in U \subset X$, $V \subset Y$, 
and $W \subset Z$.

We first define a binary Lagrangian functional
\begin{equation*}
\mathcal{L}: U \times V \to \mathbb{R}
\end{equation*}
that gives a unary functional when we substitute the solution of the implicit
function,
\begin{equation*}
\mathcal{L}(x) = \mathcal{L}(x, f(x))
\end{equation*}
with the Fr\'echet derivative
\begin{equation*}
\frac{\delta \mathcal{L} }{\delta x} (x) = 
J^{\dagger}_{g \circ f}(x)(\boldsymbol{\alpha}).
\end{equation*}

Next we introduce a mapping $\Lambda: W \rightarrow \mathbb{R}$ that
preserves the kernel of the constraint function, $\Lambda \circ c(x, y) = 0$ 
whenever $c(x, y) = 0$.  Composing this mapping with the constraint function
gives the constraint Lagrangian functional,
\begin{equation*}
\mathcal{Q} = \Lambda \circ c : U \times V \rightarrow \mathbb{R},
\end{equation*}
with $\mathcal{Q}(x, f(x)) = 0$ for all $x \in U$.

Together these two functionals define an augmented Lagrangian functional,
\begin{equation*}
\mathcal{J} = \mathcal{L} + \mathcal{Q},
\end{equation*}
along with the corresponding unary functional
\begin{align*}
\mathcal{J}(x) 
&= \mathcal{J}(x, f(x))
\\
&= \mathcal{L}(x, f(x)) + \mathcal{Q}(x, f(x))
\\
&= \mathcal{L}(x, f(x)),
\end{align*}
because the constraint functional vanishes by construction when evaluated 
at the solution to the constraint problem.

The total derivative of this augmented unary functional is given by 
\begin{align*}
\frac{\delta \mathcal J}{\delta x} (x) 
&=
\frac{\delta \mathcal J}{\delta x} (x, f(x)) 
\\
&=
\frac{\delta \mathcal J}{\partial x}(x, f(x)) 
+ \frac{\delta \mathcal J}{\delta y} (x, f(x)) \circ \frac{\delta f}{\delta x} (x).
\end{align*}

If we could engineer a map $\Lambda$ such that the \emph{adjoint system} $\delta \mathcal J / \delta y$
vanishes,
\begin{align*}
0 
&= 
\frac { \delta \mathcal{J} }{ \delta y } (x, f(x)) 
\\
&=
\frac { \delta \mathcal{L} }{ \delta y } (x, f(x))
+ \frac { \delta \mathcal{Q} }{ \delta y } (x, f(x))
\\
&=
\frac { \delta \mathcal{L} }{ \delta y } (x, f(x)) 
+ \frac{ \delta \Lambda }{ \delta c }(x, f(x)) \circ \frac{ \delta c }{ \delta y }(x, f(x))
\end{align*}
then the reverse directional derivative would reduce to
\begin{align*}
J^{\dagger}_{g \circ f}(x)(\boldsymbol{\alpha}) 
&= 
\frac{ \delta \mathcal{J} }{ \delta x }(x, f(x))
\\
&=
\frac{ \delta \mathcal{L} }{ \delta x }(x, f(x))
+ \frac{ \delta \mathcal{Q} }{ \delta x }(x, f(x))
\\
&=
\frac{ \delta \mathcal{L} }{ \delta x }(x, f(x))
+ \frac{ \delta \Lambda }{ \delta c }(x, f(x)) \circ \frac{ \delta c }{ \delta x }(x, f(x)).
\end{align*}
Unfortunately if $Y$ is infinite-dimensional then the adjoint system also becomes 
infinite-dimensional, and the general Fr\'echet derivatives will typically be too 
ungainly to implement in practice.

One important exception is when $Y$ is a \emph{Sobolev space}.  Informally a Sobolev 
space of order $k$ is an infinite-dimensional Banach space comprised of integrable, 
real-valued functions whose first $k$ derivatives are sufficiently well-defined.  What 
makes Sobolev spaces so useful is that a large class of functionals over these spaces 
can be written as integrals.

Consider for example the input space $T \subseteq \mathbb{R}$, the output space 
$S \subseteq \mathbb{R}^{N}$, and the Sobolev space $Y$ of $k$-times 
differentiable functions $\mathbf{y} : T \rightarrow S$, such as those arising 
from the solutions to $k$-th order ordinary differential equations. Any integral 
of the form
\begin{align*}
\mathcal{G}(\mathbf{y})
&=
\int_{T} \mathrm{d} t \, g \left( t, \mathbf{y}(t), \frac{ \mathrm{d} \mathbf{y} }{ \mathrm{d} t}(t), \ldots, 
\frac{ \mathrm{d}^{K} \mathbf{y} }{ \mathrm{d} t^{K}}(t) \right)
\\
&\equiv
\int_{T} \mathrm{d} t \, g \! \left( t, \mathbf{y}(t), \mathbf{y}^{(1)}(t), \ldots, \mathbf{y}^{(K)}(t) \right)
\end{align*}
defines a unary, real-valued functional $\mathcal{G} : Y \rightarrow \mathbb{R}$ whose 
Fr\'echet derivative is given by
\begin{align*}
\frac{ \delta \mathcal{G} }{ \delta \mathbf{y} }(\mathbf{y}) 
&=
\frac{ \delta }{ \delta \mathbf{y} } 
\int_{T} \mathrm{d} t \, g \! \left( t, \mathbf{y}(t), \mathbf{y}^{(1)}(t), \ldots, \mathbf{y}^{(K)}(t) \right)
\\
&=
\int_{T} \mathrm{d} t \,
\Bigg[ \hspace{14mm}
\frac{\partial g}{ \partial \mathbf{\mathbf{y}}} \!
\left(t, \mathbf{y}(t), \mathbf{y}^{(1)}(t), \ldots, \mathbf{y}^{(K)}(t) \right)
\\
&\hspace{16mm}
+ \sum_{k = 1}^{K} \frac{\partial g}{ \partial \mathbf{y}^{(k)} } \! 
\left(t, \mathbf{y}(t), \mathbf{y}^{(1)}(t), \ldots, \mathbf{y}^{(K)}(t) \right)
\cdot \frac{ \delta \mathbf{y}^{(k)} }{ \delta \mathbf{y} } 
\Bigg].
\end{align*}

Similarly if $X \subseteq \mathbb{R}^{I}$ then 
\begin{equation*}
\mathcal{G}(x, \mathbf{y})
=
\int_{T} \mathrm{d} t \, g \left(x, t, \mathbf{y}(t), \mathbf{y}^{(1)}(t), \ldots, \mathbf{y}^{(K)}(t) \right)
\end{equation*}
defines a binary, real-valued functional $\mathcal{G} : X \times Y \rightarrow \mathbb{R}$.
Consequently we can construct a large class of Lagrangian functionals and constraint functionals 
as integrals,
\begin{align*}
\mathcal{L}(x, \mathbf{y})
&=
\int_{T} \mathrm{d} t \, l \! \left(x, t, \mathbf{y}(t), \mathbf{y}^{(1)}(t), \ldots, \mathbf{y}^{(K)}(t) \right)
\\
\mathcal{Q}(x, \mathbf{y})
&=
\int_{T} \mathrm{d} t \, q \! \left(x, t, \mathbf{y}(t), \mathbf{y}^{(1)}(t), \ldots, \mathbf{y}^{(K)}(t) \right)
\end{align*}
with the corresponding augmented Lagrangian,
\begin{align*}
\mathcal{J}(x, \mathbf{y})
&=
\int_{T} \mathrm{d} t \, j \! \left(x, t, \mathbf{y}(t), \mathbf{y}^{(1)}(t), \ldots, \mathbf{y}^{(K)}(t) \right)
\\
&=
\int_{T} \mathrm{d} t \, 
\bigg[ \quad l \! \left(x, t, \mathbf{y}(t), \mathbf{y}^{(1)}(t), \ldots, \mathbf{y}^{(K)}(t) \right)
\\
& \hspace{16mm} + q \! \left(x, t, \mathbf{y}(t), \mathbf{y}^{(1)}(t), \ldots, \mathbf{y}^{(K)}(t) \right) \bigg].
\end{align*}
The choice of $l$ needs to verify the condition that
\begin{align*}
J^\dagger_{g \circ f}(x) (\boldsymbol \alpha) 
&= 
\frac{\delta}{\delta x} \mathcal L(x, f(x))  
\\
&= 
\frac{\delta}{\delta x} \int_T \ \text d t \ l \! \left(x, t, \mathbf{y}(t), \mathbf{y}^{(1)}(t), \ldots, \mathbf{y}^{(K)}(t) \right) 
\\
&= 
\int_T \mathrm d t \, \bigg[ \quad
\frac{\partial l}{\partial x}  \! \left(x, t, \mathbf{y}(t), \mathbf{y}^{(1)}(t), \ldots, \mathbf{y}^{(K)}(t) \right) 
\\
& \hspace{16mm} + \sum_{k = 1}^{K} \frac{\partial l}{ \partial \mathbf{y}^{(k)} } \! 
\left(t, \mathbf{y}(t), \mathbf{y}^{(1)}(t), \ldots, \mathbf{y}^{(K)}(t) \right)
\cdot \frac{ \delta \mathbf{y}^{(k)} }{ \delta x } \bigg].
\end{align*}
Depending on the problems, choosing $l$ can be straightforward or require a bit more work, as we will see in the examples.
The choice of $q$ needs to satisfy the condition that, for all $x \in U$,
\begin{eqnarray*}
  0 & = & \mathcal Q (x, f(x)) \\ 
    & = & \int_T \ \text d t \ q \! \left(x, t, \mathbf{y}(t), \mathbf{y}^{(1)}(t), \ldots, \mathbf{y}^{(K)}(t) \right).
\end{eqnarray*}
One straightforward strategy is to define the integrand $q$ such that it vanishes whenever evaluated at the solution, $(x, f(x))$.
We can accomplish this for example by taking $q$ to be the Sobolev inner product of the constraint function, $f$, with an auxiliary function $\lambda \in Z$,
\begin{equation*}
  q(x, f(x)) = \braket{\lambda, \  c(x, f(x))},
\end{equation*}
where $\braket{ \ , \ }$ denotes the Sobolev inner product.
When evaluated at the solution the constraint function $c$ vanishes so that the above inner product, and hence the integrand $q$, also vanish.


When using functionals of this form the reverse directional derivative becomes
\begin{align*}
J^{\dagger}_{g \circ f}(x)(\boldsymbol{\alpha}) 
&= 
\frac{ \delta \mathcal{J} }{ \delta x }(x, f(x))
\\
&=
\int_{T} \mathrm{d} t \,
\Bigg[ \quad
\frac{\partial j}{ \partial x} \!
\left(x, t, \mathbf{y}(t), \mathbf{y}^{(1)}(t), \ldots, \mathbf{y}^{(K)}(t) \right)
\\
&\hspace{16mm}
+ \frac{\partial j}{ \partial \mathbf{\mathbf{y}}} \!
\left(x, t, \mathbf{y}(t), \mathbf{y}^{(1)}(t), \ldots, \mathbf{y}^{(K)}(t) \right)
\cdot \frac{ \mathrm{d} \mathbf{y} }{ \mathrm{d} x } (t)
\\
&\hspace{16mm}
+ \sum_{k = 1}^{K} \frac{\partial j}{ \partial \mathbf{y}^{(k)} } \! 
\left(x, t, \mathbf{y}(t), \mathbf{y}^{(1)}(t), \ldots, \mathbf{y}^{(K)}(t) \right)
\cdot \frac{ \delta \mathbf{y}^{(k)} }{ \delta \mathbf{y} } 
\cdot \frac{ \mathrm{d} \mathbf{y} }{ \mathrm{d} x } (t)
\Bigg]
\\
&=
\int_{T} \mathrm{d} t \,
\Bigg[ \quad
\frac{\partial j}{ \partial x} \!
\left(x, t, \mathbf{y}(t), \mathbf{y}^{(1)}(t), \ldots, \mathbf{y}^{(K)}(t) \right)
\\
&\hspace{16mm}
+ \frac{\partial j}{ \partial \mathbf{\mathbf{y}}} \!
\left(x, t, \mathbf{y}(t), \mathbf{y}^{(1)}(t), \ldots, \mathbf{y}^{(K)}(t) \right)
\cdot \frac{ \mathrm{d} \mathbf{y} }{ \mathrm{d} x } (t)
\\
&\hspace{16mm}
+ \sum_{k = 1}^{K} \frac{\partial j}{ \partial \mathbf{y}^{(k)} } \! 
\left(x, t, \mathbf{y}(t), \mathbf{y}^{(1)}(t), \ldots, \mathbf{y}^{(K)}(t) \right)
\cdot \frac{ \mathrm{d} \mathbf{y}^{(k)} }{ \mathrm{d} x } (t)
\Bigg].
\end{align*}
Because the elements of the Sobalev space are continuous we can simplify the terms in 
the last line by repeated application of Stokes' Theorem,
\begin{align*}
\int_{T} \mathrm{d} t \,
\frac{\partial j}{ \partial \mathbf{y}^{(k)} }
\cdot \frac{ \mathrm{d} \mathbf{y}^{(k)} }{ \mathrm{d} x }
&= \quad
\sum_{k' = 0}^{k} (-1)^{k'} \left[ \frac{ \mathrm{d}^{k'} }{ \mathrm{d} t^{k'} }
\left( \frac{ \partial j }{ \partial \mathbf{y}^{(k)} } \right)
\frac{ \mathrm{d}^{k - k' - 1} }{ \mathrm{d} t^{k - k' - 1} } 
\left( \frac{ \mathrm{d} \mathbf{y} }{ \mathrm{d} x } \right) \right]_{\partial T}
\\
& \quad
+ (-1)^{k} \int_{T} \mathrm{d} t \, \frac{ \mathrm{d}^{k} }{ \mathrm{d} t^{k} } 
\left( \frac{ \partial j }{ \partial \mathbf{y}^{(k)} } \right) 
\cdot \frac{ \mathrm{d} \mathbf{y} }{ \mathrm{d} x },
\end{align*}
where we have dropped the arguments to ease the notational burden.  Substituting 
this into the reverse directional derivative gives
\begin{align*}
J^{\dagger}_{g \circ f}(x)(\boldsymbol{\alpha}) 
&= \quad
\int_{T} \mathrm{d} t \,
\frac{\partial j}{ \partial x}
\\
& \quad+
\sum_{k = 1}^{K} \sum_{k' = 0}^{k} (-1)^{k'} 
\left[ \frac{ \mathrm{d}^{k'} }{ \mathrm{d} t^{k'} }
\left( \frac{ \partial j }{ \partial \mathbf{y}^{(k)} } \right)
\frac{ \mathrm{d}^{k - k' - 1} }{ \mathrm{d} t^{k - k' - 1} } 
\left( \frac{ \mathrm{d} \mathbf{y} }{ \mathrm{d} x } \right) \right]_{\partial T}
\\
& \quad+
\int_{T} \mathrm{d} t \,
\left[ \frac{\partial j}{ \partial \mathbf{\mathbf{y}}}
+ \sum_{k = 1}^{K}
(-1)^{k} \frac{ \mathrm{d}^{k} }{ \mathrm{d} t^{k} } 
\left( \frac{ \partial j }{ \partial \mathbf{y}^{(k)} } \right) \right]
\cdot \frac{ \mathrm{d} \mathbf{y} }{ \mathrm{d} x }.
\end{align*}

If we can engineer a $q$ such that $j$ solves the \emph{differential adjoint system}
\begin{align*}
0
&=
\sum_{k = 1}^{K} \sum_{k' = 0}^{k} (-1)^{k'} 
\left[ \frac{ \mathrm{d}^{k'} }{ \mathrm{d} t^{k'} }
\left( \frac{ \partial j }{ \partial \mathbf{y}^{(k)} } \right)
\frac{ \mathrm{d}^{k - k' - 1} }{ \mathrm{d} t^{k - k' - 1} } 
\left( \frac{ \mathrm{d} \mathbf{y} }{ \mathrm{d} x } \right) \right]_{\partial T}
\\
0
&=
\frac{\partial j}{ \partial \mathbf{\mathbf{y}}}
+ \sum_{k = 1}^{K}
(-1)^{k} \frac{ \mathrm{d}^{k} }{ \mathrm{d} t^{k} } 
\left( \frac{ \partial j }{ \partial \mathbf{y}^{(k)} } \right)
\end{align*}
then the reverse directional derivative reduces to the manageable form
\begin{align*}
J^{\dagger}_{g \circ f}(x)(\boldsymbol{\alpha}) 
&= 
\int_{T} \mathrm{d} t \, \frac{\partial j}{ \partial x} 
\left(x, t, \mathbf{y}(t), \mathbf{y}^{(1)}(t), \ldots, \mathbf{y}^{(K)}(t) \right)
\\
&= \int_{T} \mathrm{d} t \, \bigg[ \quad
\frac{\partial l}{ \partial x} \left(x, t, \mathbf{y}(t), \mathbf{y}^{(1)}(t), \ldots, \mathbf{y}^{(K)}(t) \right)
\\
& \hspace{15mm} + 
\frac{\partial q}{ \partial x} \left(x, t, \mathbf{y}(t), \mathbf{y}^{(1)}(t), \ldots, \mathbf{y}^{(K)}(t) \right)
\bigg].
\end{align*}

For different choices of $Y$ the form of the functionals, their Fr\'echet derivatives,
and the resulting differential adjoint system will vary, but the basic procedure of 
the adjoint method remains the same.  By considering more sophisticated Sobelev spaces
this general adjoint method can be applied to partial differential equations and other
more sophisticated infinite-dimensional implicit systems.

\subsection{Demonstrations} \label{sec:inf-demo}

The particular differential adjoint system we have derived in Section~\ref{sec:adjoint-inf} 
is immediately applicable to implicit systems defined by ordinary differential and 
algebraic differential equations. In this section we demonstrate that application on 
two such systems.

\subsubsection{Ordinary Differential Equations.}

Consider a time interval $T = [0, \tau] \subset \mathbb{R}$ and the $N$-dimensional
trajectories that map each time point to an $N$-dimensional state, 
$\mathbf{y} : T \rightarrow \mathbb{R}^{N}$.  The space of trajectories that are at
least once-differentiable forms a first-order Sobalev space, $Y$.

A linear system of first-order, ordinary differential equations
\begin{equation*}
\frac{ \mathrm{d} \mathbf{y} }{ \mathrm{d} t } = \mathbf{r}(x, \mathbf{y}, t),
\end{equation*}
along with an initial condition $\mathbf{y}(0) = \mathbf{u}(x)$, defines a separate 
constraint for the trajectory behavior at each $t \in T$,
\begin{align*}
c(x, \mathbf{y})(t) 
&= 
\frac{ \mathrm{d} \mathbf{y} }{ \mathrm{d} t } (t) - \mathbf{r} (x, \mathbf{y})(t)
\\
c(x, \mathbf{y})(0) 
&= 
\mathbf{y}(0) - \mathbf{u}(x).
\end{align*}
Collecting all of these constraints together defines an infinite-dimensional constraint 
function $c : X \times Y \to Z$ where $Z$ is also the space of differentiable functions 
that map from $T$ to $\mathbb R^N$.

The summary function $g : \mathbf{y} \mapsto \mathbf{y}(\tau)$ projects
infinite-dimensional trajectories down to their $N$-dimensional final states 
so that the composition $g \circ f$ maps inputs $x \in X$ to a final state at
time $t = \tau$.  In order to implement this map into a reverse mode automatic
differentiation library we need to be able to implement the reverse directional 
derivative $J_{g \circ f}(x)(\boldsymbol{\alpha})$.

To implement the adjoint method we need a Lagrangian functional that satisfies
\begin{equation*}
\frac{ \mathrm{d} \mathcal{L} }{ \mathrm{d} x_{i} }(x, f(x))
=  \left( \frac{ \mathrm{d} \mathbf{y} }{ \mathrm{d} x_{i} }(T) \right)^T \cdot \boldsymbol \alpha.
\end{equation*}
Integrating the defining differential equation gives an integral functional that
provides a particularly useful option,
\begin{equation*}
\mathcal{L}(x, \mathbf{y})
= \mathbf{u}^T(x) \cdot \boldsymbol \alpha 
+ \int_{0}^{\tau} \mathrm{d} t \ \mathbf{r}^T(x, \mathbf{y}, t) \cdot \boldsymbol \alpha.
\end{equation*}

Next we need to construct a functional $\mathcal Q$ that vanishes when when evaluated at 
the implicit solution.  Integrating over the constraint function gives
\begin{equation*}
\mathcal{Q}(x, \mathbf{y})
=
\bigg[ \mathbf{y}(0) - \mathbf{u}(x) \bigg]^T \cdot \boldsymbol{\mu}
+ \int_{0}^{T} \mathrm{d} t \, 
\left[ \frac{ \mathrm{d} \mathbf{y} }{ \mathrm{d} t }(t) - \mathbf{r}(x, \mathbf{y}, t) \right]^T 
\cdot \boldsymbol{\lambda}(t),
\end{equation*}
for any function $\boldsymbol{\lambda}: \mathbb R \to \mathbb R^N$ and constant 
$\boldsymbol{\mu} \in \mathbb{R}^{N}$.

Substituting these integral functionals into the differential adjoint system that
we derived in Section~\ref{sec:adjoint-inf} yields the system
\begin{align*}
0 &= 
\left( \frac{ \mathrm{d} \mathbf{y} }{ \mathrm{d} x } (0) \right)^T 
\cdot \left( \boldsymbol{\mu} - \boldsymbol{\lambda}(0) \right) 
+ 
\left( \frac{ \mathrm{d} \mathbf{y} }{ \mathrm{d} x } (\tau) \right)^T 
\cdot \boldsymbol{\lambda}(\tau)
\\
0 &=
\left( \frac{ \partial \mathbf{r} }{ \partial \mathbf{y} } (x, \mathbf{y}, t) \right)^T 
\cdot \left( \boldsymbol{\alpha} - \boldsymbol{\lambda}(t) \right)
- \frac{ \mathrm{d} \boldsymbol{\lambda} }{ \mathrm{d} t }(t).
\end{align*}
The boundary terms vanish if we take $\boldsymbol{\mu} = \boldsymbol{\lambda}(0)$
and $\boldsymbol{\lambda}(\tau) = 0$ while the differential term vanishes for the
$\boldsymbol{\lambda}(t)$ given by integrating $\boldsymbol{\lambda}(\tau) = 0$ 
backwards from $t = \tau$ to $t = 0$.  In other words the differential adjoint 
system defines a linear, first-order ordinary differential equation that reverses 
time relative to our initial ordinary differential equation.

Once we've solved for $\boldsymbol{\lambda}(t)$ the reverse directional 
derivative is given by the partial derivatives of the augmented Lagrangian 
with respect to $x$,
\begin{equation*}
J^{\dagger}_{g \circ f}(x)(\boldsymbol{\alpha})
=
\left( \frac{ \partial \mathbf{u} }{ \partial x } (x) \right)^T 
\cdot \left( \boldsymbol{\alpha} - \boldsymbol{\lambda}(0) \right)
+ \int_{0}^{T} \mathrm{d} t \,
\left (\frac{ \partial \mathbf{r} }{ \partial x } (x, \mathbf{y}, t) \right)^T 
\cdot \left( \boldsymbol{\alpha} - \boldsymbol{\lambda}(t) \right).
\end{equation*}
While we do have to solve both the nominal and adjoint differential equations,
these solves require evaluating only finite-dimensional derivatives.  We have 
completely avoided any infinite-dimensional Fr\'{e}chet derivatives.

\subsubsection{Differential Algebraic Equations.}

Introducing an algebraic constraint to the previous systems defines a
differentiable algebraic system, or DAE. A DAE might, for example, impose
the constraint that the component states at each time sum to one so that 
the states can model how the allocation of a conserved quantity evolves
over time.

To simplify the derivation we start by decomposing the trajectories
$\mathbf{y}(t) \in \mathbb{R}^{N}$ into a differential component,
$\mathbf{y}^{d} \in \mathbb{R}^{D}$, and an algebraic component,
$\mathbf{y}^{a} \in \mathbb{R}^{A}$, with $N = D + A$.  A differential
algebraic constraint function can similarly be decomposed into
a differential constraint function,
\begin{equation*}
c^d(x, \mathbf y, \dot{\mathbf{y}}, t) \in \mathbb{R}^{D},
\end{equation*}
where $\dot{\mathbf{y}}$ is shorthand for $\mathrm{d} \mathbf{y} / \mathrm{d} t$,
and an algebraic constraint function,
\begin{equation*}
c^a (x, \mathbf{y}, t) \in \mathbb{R}^{A}.
\end{equation*}
with $\mathbf{c} = (c^d, c^a)^{T}$.

If the differential constraint is given by a linear, first-order differential 
equation then the differential constraint function becomes
\begin{align*}
c^d(x, \mathbf y, \dot{\mathbf{y}}, t) &= \dot{\mathbf y}^d - \mathbf r^d(x, \mathbf y, t)
\\
c^{d}(x, \mathbf{y}, \dot{\mathbf{y}}, 0) &= \mathbf y (0) - \mathbf u (x).
\end{align*}

When the constraints are consistent this differential algebraic system implicitly 
defines a map from inputs $x \in X$ to $N$-state trajectories, 
$\mathbf{y} \in T \times \mathbb{R}^{N}$.  We can also write this as a map from 
inputs and times to states,
\begin{equation*}
f : X \times T \rightarrow \mathbb{R}^{N},
\end{equation*}
such that $c^d(x, \mathbf{f}(x), \dot{\mathbf{f}}(x), t) = 0$ and 
$c^a (x, \mathbf{f}(x), t) = 0$.

As in the previous example we will consider a summary function that projects the 
infinite-dimensional trajectories down to their $N$-dimensional final states,
$g : \mathbf{y} \mapsto \mathbf{y}(\tau)$, so that the composition $g \circ f$ 
maps inputs $x \in X$ to a final state at time $t = \tau$.  Our goal is then to
evaluate the finite-dimensional reverse directional derivative 
$J_{g \circ f}(x)(\boldsymbol{\alpha})$.

As with ordinary differential systems integrating the trajectory provides an
appropriate integral functional,
\begin{equation*}
\mathcal{L}(x, \mathbf{y})
= \mathbf{u}^T(x) \cdot \boldsymbol \alpha 
+ \int_{0}^{\tau} \mathrm{d} t \ \dot{\mathbf{y}}^T (x, t) \cdot \boldsymbol \alpha
\end{equation*}
that satisfies
\begin{equation*}
\frac{ \mathrm{d} \mathcal{L} }{ \mathrm{d} x }(x, \mathbf{f}(x))
= \left (\frac{ \mathrm{d} \mathbf{f} }{ \mathrm{d} x}(x, \tau) \right)^T \cdot \boldsymbol \alpha.
\end{equation*}
Unlike in the ordinary differential case, however, the differential algebraic system
does not immediately provide an analytical expression for $\dot{\mathbf{y}}(x)$.

The first-order linear differential equation does provide the derivative of the
differential component,
\begin{equation*}
\dot{\mathbf{y}}^{d} = \mathbf r^d.
\end{equation*}
In order to obtain the derivative of the algebraic component we have to differentiate
the algebraic constraint,
\begin{eqnarray*}
  0 & = & \frac{\mathrm d}{\mathrm d t} c^a (x, \mathbf y, t)  \\
    & = & \frac{\partial c^a}{\partial \mathbf y^d} \dot{\mathbf y}^d (x, t) + \frac{\partial c^a}{\partial \mathbf y^a} \dot{\mathbf y}^a (x, t)
    + \frac{\partial c^a}{\partial t} (x, t),
\end{eqnarray*}
or
\begin{equation*}
\dot{\mathbf y}^a 
= -
\left [  \frac{\partial c^a}{\partial \mathbf y^a} \right ]^{-1} 
\left( \frac{\partial c^a}{\partial \mathbf y^d} \dot{\mathbf y}^d + \frac{\partial c^a}{\partial t} \right),
\end{equation*}
where we assume $\partial c^a / \partial \mathbf y^a$ is square invertible.

The constraint functional $\mathcal Q$ is identical to that from the ordinary differential 
equation system,
\begin{equation*}
\mathcal{C}(x, \mathbf{y})
=
\bigg[ \mathbf{y}(0) - \mathbf{u}(x) \bigg]^T \cdot \boldsymbol \mu
+ 
\int_{0}^{\tau} \mathrm{d} t \, \mathbf c(x, \mathbf y, t)^T \cdot \boldsymbol{\lambda}(t).
\end{equation*}

Plugging $\mathcal{J} = \mathcal{L} + \mathcal{Q}$ into the results of 
Section~\ref{sec:adjoint-inf} gives the system
\begin{align*}
0
&=
\left [\frac{\text d \mathbf y}{\text d x_i}(0) - \frac{\text d \mathbf u}{\text d x_i} (x) \right]^T 
\cdot \boldsymbol \mu
+ \left. \left( \frac{\text d \mathbf y}{\text d x_i} \right)^T 
\cdot \boldsymbol \lambda_{D0}(t) \right |^\tau_0 
\\
0
&=
\left( \frac{\partial \mathbf r}{\partial \mathbf y} (x, \boldsymbol y, t) \right)^T \cdot \boldsymbol \alpha
    +  \left(\frac{\partial \mathbf c}{\partial \mathbf y} (x, \boldsymbol y, t)\right)^T \cdot \boldsymbol \lambda(t) - \boldsymbol{\dot \lambda_\text{\emph D0}}(t)
\end{align*}
where
\begin{align*}
\boldsymbol \lambda_{D0} (t)
&=
\left(\frac{\partial \mathbf c}{\partial \mathbf{\dot y}} (x, \mathbf y, t) \right)^T \cdot \boldsymbol \lambda (t) 
\\
&=
\begin{bmatrix}  
    \frac{\partial \mathbf c^d}{\partial \mathbf{\dot y_\text{\emph d}}} & 
    \frac{\partial \mathbf c^d}{\partial \mathbf{\dot y_\text{\emph a}}} \\ 
    \frac{\partial \mathbf c^a}{\partial \mathbf{\dot y_\text{\emph d}}} & 
    \frac{\partial \mathbf c^a}{\partial \mathbf{\dot y_\text{\emph a}}}
\end{bmatrix}
\cdot \boldsymbol \lambda (t) 
\\
&=
\begin{bmatrix}
    \mathbf I_{D \times D} & \mathbf 0_{D \times A} \\
    \mathbf 0_{A \times D} & \mathbf 0_{A \times A}
\end{bmatrix}
\cdot 
\boldsymbol \lambda (t) 
\\
&= 
\begin{bmatrix}
  \boldsymbol \lambda_{1:D} (t) \\ \mathbf 0_{1 \times A}
\end{bmatrix}.
\end{align*}

Decomposing each term in the differential adjoint system into algebraic and a differential components
gives
\begin{eqnarray*}
  \begin{bmatrix}
    \left (\frac{\partial \mathbf r}{\partial \mathbf y_\text{\emph d}} (x, \mathbf y, t) \right)^T \cdot \boldsymbol \alpha
    + \left(\frac{\partial \mathbf c_d}{\partial \mathbf y_d} (x, \mathbf y, t) \right)^T \cdot \boldsymbol \lambda_d (t)
    + \left(\frac{\partial \mathbf c_a}{\partial \mathbf y_d} (x, \mathbf y, t) \right)^T \cdot \boldsymbol \lambda_a (t) - \boldsymbol{\dot \lambda_\text{\emph D0}}(t) \\
\left(\frac{\partial \mathbf r}{\partial \mathbf y_a} (x, \mathbf y, t) \right)^T \cdot \boldsymbol \alpha
    + \left( \frac{\partial \mathbf c_d}{\partial \mathbf y_a} (x, \mathbf y, t) \right)^T \cdot \boldsymbol \lambda_d (t)
    + \left(\frac{\partial \mathbf c_a}{\partial \mathbf y_a} (x, \mathbf y, t)\right)^T \cdot \boldsymbol \lambda_a (t)
   \end{bmatrix}
   = 
   \begin{bmatrix}
   \mathbf 0_D \\ \mathbf 0_A
  \end{bmatrix}.
\end{eqnarray*}
which defines an adjoint DAE

To ensure that the boundary term vanish we need to set $\boldsymbol \mu = \boldsymbol \lambda_{D0}(0)$
and $\boldsymbol \lambda_{D0} (\tau) = \mathbf 0_N$.  The algebraic component of $\boldsymbol{\lambda}$ 
at $t = \tau$ is also given by
\begin{align*}
\mathbf 0_A &= 
\left( \frac{\partial \mathbf r}{\partial \mathbf y_a} (x, \mathbf y, \tau) \right)^T 
\cdot \boldsymbol \alpha
+ 0
+ \left(\frac{\partial \mathbf c_a}{\partial \mathbf y_a} (x, \mathbf y, \tau) \right)^T 
\cdot \boldsymbol \lambda_a (\tau)
\\
\boldsymbol \lambda_a (\tau) 
&= 
- \left [ \left( \frac{\partial \mathbf c_a}{\partial \mathbf y_a} (x, \mathbf y, \tau) \right) \right ]^{-1} 
\cdot \left(\frac{\partial \mathbf r}{\partial \mathbf y_a} (x, \mathbf y, \tau) \right)^T \cdot \boldsymbol \alpha,
\end{align*}
where we recall our assumption that $\partial \mathbf c_a / \partial \mathbf y_a$ is square invertible.

Now we can ensure that the differential term vanishes by integrating this adjoint DAE backwards
from the terminal condition 
$\boldsymbol{\lambda}(\tau) = (\boldsymbol{\lambda}^{d}(\tau) , \boldsymbol{\lambda}^{a}(\tau) )^{T}$
at $t = \tau$ to an initial condition at $t = 0$.  

Once we have solved for $\boldsymbol{\lambda}(t)$ the reverse directional derivative reduces to
\begin{equation*}
J^{\dagger}_{g \circ f}(x)(\boldsymbol{\alpha}) 
= 
- \left( \frac{\text d \mathbf u}{\text d x_i} \right)^T \cdot \boldsymbol \lambda_{D0}(0) 
+ 
\int_0^\tau \text d t \ 
\left( 
\frac{\partial \mathbf r}{\partial x_i} (x, \boldsymbol y, t)
+ \boldsymbol \lambda^T \frac{\partial \mathbf c}{\partial x_i} (x, \boldsymbol y, t) 
\right)^T \cdot \boldsymbol \lambda(t).
\end{equation*}

\section{Discussion}
The implicit function theorem allows us to construct an 
expression for the directional derivatives of an implicit function as a composition 
of Fr\'echet derivatives.  The adjoint method computes these directional derivatives 
directly without evaluating any of the intermediate terms.

When the output of the implicit function is finite-dimensional these differential 
operators can be implemented with linear algebra, although care and experience is 
required to ensure that the linear algebra operations are as efficient as possible.
While the adjoint method yields the same result, it naturally incorporates any 
available structure in the implicit system so that optimal performance can be achieved 
automatically as we saw in our discussion on difference equations (Section~\ref{sec:de}).

If the output of the implicit function is infinite-dimensional then the component
Fr\'echet derivatives that make up the directional derivatives of the implicit function
can no longer be evaluated directly, making the composition intractable.
Implementing the adjoint method also requires Fr\'echet derivatives which in general 
we cannot evaluate.
In the important special case where the output of the implicit function falls into a 
Sobolev space, however, we can engineer the augmented Lagrangian so that the Fr\'echet 
derivatives reduce to tractable functional derivatives.
This is notably the strategy we deploy when in the case of ODEs and DAEs
(Section~\ref{sec:inf-demo}).


While the adjoint method is more generally applicable it is not as systematic as
the implicit function theorem method.  The practicality and performance of the 
method depends on the choice of Lagrangian and constraint functionals.  Engineering
performant functionals, let alone valid functionals at all, is by no means trivial.
Fortunately in many problems the structure of the implicit system guides the design.

Beyond the implicit function theorem and adjoint methods, we may use the trace method which automatically differentiates through the trace of a numerical solver.  
In most cases this approach leads to computationally expensive and memory intensive algorithms. 

For finite-dimensional systems we could also construct a ``forward method'' that computes the Jacobian $J = J_{g \circ f}$ before evaluating its action on an input sensitivity or adjoint to form the wanted directional derivatives.  As we discussed in Section~\ref{sec:trace}, however, fully computing $J$ first is always less efficient that the iterative evaluation of the directional derivative; see also \cite{Gaebler:2021}.

Although not as general, we can also construct a forward method that fully computes the Jacobian $J = J_{{g} \circ J_{f}}$ for certain infinite-dimensional problems.
This approach notably applies to certain classes of ODEs and DAEs (Appendix~\ref{app:forward}).
In these cases the computational trade-offs between the forward method and the adjoint method is more nuanced; which method is more efficient depends on the specific of the problem.  
For example \cite{Rackauckas:2018} compares the forward and adjoint approaches to implementing
automatic differentiation for ODEs.  For small ODE systems the overhead cost associated with solving the adjoint system can make the method relatively slow, but as the size of the system and the dimension of $x$ increases the adjoint method benefits from superior scalability.  See also \cite{Hindmarsh:2020, Betancourt:2020} for additional scaling discussions.

\section{Acknowledgment}

We thank David Childers, David Kent, and Dalton AR Sakthivadivel for helpful discussion.

\appendix

\section{Infinite-Dimensional Forward Method} \label{app:forward}

As in Section \ref{sec:infinite} consider finite-dimensional real input, $X = \mathbb R^I$,
an infinite-dimensional output space $Y$, and the summary function
$g : Y \rightarrow \mathbb{R}^{J}$.  The forward method explicitly computes the matrix 
representation of the full Jacobian, $\mathbf J_{g \circ f}$, before contracting this matrix 
with a sensitivity or adjoint vector to implement directional derivatives for automatic 
differentiation.

The implicit function theorem prescribes a composite expression for $\mathbf J_{g \circ f}$, 
but in general we cannot evaluate the intermediate Fr\'echet derivatives. In certain special 
cases, however, we can bypass the implicit function theorem and evaluate $\mathbf J_{g \circ f}$ 
directly.

As with the adjoint method, we focus on the special case where $Y$ is a Sobolev space. 
Theoretically we can reduce a key Fr\'echet derivative to a functional derivative,
which we can evaluate by solving a \textit{forward differential system}.
In practice our ability to construct such a system \textit{and} solve it depends on the 
specifics of the problem.

Let $Y$ be the order $K$ Sobolev space of functions $T \subset \mathbb{R} \rightarrow \mathbb{R}^{N}$,
and suppose that we can construct a functional
\begin{eqnarray*}
\mathcal{P} : & X \times Y  & \rightarrow \mathbb{R}^{J} \\
                & x, \mathbf{y} & \mapsto \mathcal{P}(x, \mathbf{y}),
\end{eqnarray*}
which satisfies
\begin{equation*}
  \frac{\delta \mathcal P}{\delta x} (x) = J_{g \circ f}(x).
\end{equation*}
In addition, assume there exists such a functional which takes the form of an integral,
\begin{equation*}
  \mathcal{P}(x, \mathbf{y}) = \int_{T} dt \  p(x, t, \mathbf{y}(t), ... ).
\end{equation*}
For example if our summary function $g$ is a Sobolev inner product with the implicit function,
\begin{eqnarray*}
  (g \circ f)(x) & = & \braket{\gamma, \ f(x)}  \\
                 & = & \int dt\  \gamma(t) \, f(x, t) \\
                 & = & \int dt \  \boldsymbol{\gamma}^{T}(t) \cdot \mathbf{y}(x, t)
\end{eqnarray*}
then we can take
\begin{equation*}
  p = \boldsymbol{\gamma}^{T}(t) \cdot \mathbf{y}(x, t),
\end{equation*}
and obtain a satisfactory functional $\mathcal P$.

Taking the derivative with respect to $x$ gives,
\begin{eqnarray*}
  \frac{\delta \mathcal P}{\delta x} & = & \int_T \text d t \ \frac{\text d p}{\text d x} \left (x, t, \mathbf y(t), \mathbf y^{(1)}(t), \cdots, \mathbf y^{(K)}(t) \right)  \\
    & = & \int_T \text d t \ \frac{\partial p}{\partial x} + \sum_{k = 0}^K \frac{\partial p}{\partial \mathbf y^{(k)}} \frac{\text d \mathbf y^{(k)}}{\text d x},
\end{eqnarray*}
where crucially the Fr\'echet derivative reduces to a functional derivative.

In order to evaluate this integral we need to evaluate the derivatives of the implicit function, $\text d \mathbf y^{(k)} /\text d x$, at all times $t \in T$.  
In theory we could achieve this by fully constructing the first-order Fr\'echet derivative from the implicit function theorem, repeatedly differentiating it, and then evaluating all of those Fr\'echet derivatives at each time $t$.  

The need to evaluate Fr\'echet derivatives, however, makes this approach infeasible in practice.  A more viable alternative is to evaluate the derivatives only at specific times,
where they reduce to manageable finite-dimensional objects.  

By definition our constraint function defines a map $c : X \times Y \rightarrow Z$ where $Y$ and $Z$ 
are both the space of functions which map from $T$ to $\mathbb{R}^{N}$.  In this case the constraint 
function can equivalently be defined as a collection of maps $X \times Y \rightarrow \mathbb{R}^{N}$ 
for each $t \in T$.  Denoting these maps as $c(x, y, t)$ the implicit function $f: X \rightarrow Y$ is
defined by the system of constraints,
\begin{equation*}
c(x, f(x), t) = 0, \, \forall \, t \in T.
\end{equation*}

Fixing $t$ and then differentiating with respect to the input $x$ gives
\begin{eqnarray*}
  0 & = & \frac{\partial c}{\partial x} (t) + \sum_{k = 0}^K \frac{\partial c}{\partial \mathbf y^{(k)}} \frac{\text d \mathbf y^{(k)}}{\text d x}(t)  \\
  & = & \frac{\partial c}{\partial x} (t) + \sum_{k = 0}^K \frac{\partial c}{\partial \mathbf y^{(k)}} \left( \frac{\text d \mathbf y}{\text d x} \right)^{(k)}(t).
\end{eqnarray*}
%
This \emph{forward differential system} implicitly defines the derivative evaluations at each $t$ as 
a differential equation.  Once we've solved for $\mathbf{y}(t)$ we can, at least in theory, solve this 
forward differential system for each $\text d \mathbf{y}^{(k)} / \mathrm{d} x \, (x, f(x), t)$ and then evaluate
\begin{equation*}
  \int_T \text d t \ \sum_{k = 0}^K \frac{\partial p}{\partial \mathbf y^{(k)}} \frac{\text d \mathbf y^{(k)}}{\text d x} = \sum_{k = 0}^K \int_T \text d t \ \frac{\partial p}{\partial \mathbf y^{(k)}} \frac{\text d \mathbf y^{(k)}}{\text d x},
\end{equation*}
as a sum of one-dimensional numerical quadratures.

If $c(x, y, t)$ is a linear a function of $\text d y/ \text d t$ and does not depend on higher-order derivatives,
\begin{equation*}
  c(\mathbf y, x, t) = \frac{\text d \mathbf y}{\text d t} (t) - f(x, \mathbf y, t) = 0,
\end{equation*}
then the forward differential system becomes particularly manageable.  In particular the forward differential 
system is also linear in the first-order derivative with respect to $t$,
\begin{equation*}
  \frac{\text d c}{\text d x} (\mathbf y, x, t) = \frac{\text d}{\text d t} \frac{\text d \mathbf y}{\text d x} (t) - \frac{\partial f}{\partial x} (x, \mathbf y, t) - \frac{\partial f}{\partial \mathbf y} \frac{\text d \mathbf y}{\text d t} (t) = 0.
\end{equation*}
In the absence of such a linearity we need to solve for the evaluations of the trajectory $\mathbf{y}$, the first-order derivative $\text d \mathbf y / \text d x$, and the higher-order derivatives of $\mathbf{y}$ at the given $x$ and \emph{every} $t$ needed for the numerical quadratures.

For a demonstration of this forward approach on certain ODEs and DAEs see \citep{Hindmarsh:2020}.

Finally similar to the Sobolev adjoint method (Section~\ref{sec:adjoint-inf}) the above derivation generalizes immediately to constrained systems over Sobolev spaces of functions $T \subset \mathbb R^M \to \mathbb R^N$, with $M > 1$.
Here instead of an ordinary differential forward system we recover a partial differential forward system, and the Jacobian is recovered as a sum of multidimensional integrals instead of one-dimensional integrals.

\bibliography{ad_implicit}
\bibliographystyle{imsart-nameyear}

\end{document}